\documentclass[12pt]{article}
\usepackage{epsfig}

\begin{document}

\title{QED \\
and ortho-para- positronium \\
mass difference}
\author{G.V.Efimov  
\footnote{
E-mail: efimovg@theor.jinr.ru,
FAX: 7(49621)65084
}
\vspace*{0.2\baselineskip}\\
 \itshape Bogoliubov Laboratory of Theoretical Physics,\\
 \itshape Joint Institute for Nuclear Research,\\
{\it 141980 Dubna, Russia}\vspace*{0.2\baselineskip} }
\maketitle

\begin{abstract}

Bound state problem in the relativistic QED is investigated by the
functional integral methods. The ortho- para- positron mass
difference is calculated.  Contribution of the "nonphysical"$~$ time
variable turned out to be important and leads to the nonanalytic
dependence of the bound state mass of the order
$\alpha^{{2\over3}}$. It is shown that the relativistic and
non-relativistic QED gives different results for this mass shift. In
addition so-called abnormal states as "time excitations" arise.
Sequential application of relativistic QED to bound state problem is
in contradiction with real ortho- and para- positronium bound
states.

The conclusion:  the relativistic QED is not suited to describe real
bound states correctly.
\end{abstract}

\vspace{0.5cm}
\noindent
Pacs Numbers: 13.20.-v, 13.20.Fc, 13.20.He, 24.85.+p

\newpage

\section{Introduction.}

We believe that the relativistic quantum electrodynamics (QED) is a
uniquely correct universal theory giving an exhaustive description
of all interactions between electrons and photons including possible
bound states like positronium. Only our inability to calculate
something out of perturbation method does not permit us to obtain
all the desired details. Earlier, some scientists considered that
QED should have its own applicability region. A short review of the
history and the development of quantum field theory is done in
\cite{Wein}. Supporting these doubts  we will show in this paper
that the sequential use of the standard QED does not give a correct
description of the positronium spectra, namely, the ortho- para-
positronium mass difference.

First of all let us realize what is the status of bound states in
non-relativistic quantum mechanics (QM) and relativistic quantum
field theory (QFT). In what follows, we restrict ourselves to the
discussion of positronium in QED. The difference between QM and QFT
is shown in Table I. Let us give some comments.

The total Hamiltonian $H=H_0+gH_I$ can be constructed in QM and QFT.
However, in QM  $H$ is a well defined operator, so that the
non-relativistic Schr\"{o}dinger equation is mathematically correct
and time development of a quantum system can be described. Solutions
of the Schr\"{o}dinger equation contain both free and bound states.
One can remark that in QM a bound state (positronium) is created by
real particles (electron-positron), i.e. constituent particles are
on mass shell but are no virtual particles.

In QFT the Fock space ${\cal F}$ is defined by the noninteracting
free Hamiltonian $H_0$ and contains the free particles only.
However, $H_I$ is not defined on ${\cal F}$. As a result, the bound
state as an eigenvalue problem of the relativistic Schr\"{o}dinger
equation on the Fock space cannot be formulated mathematically in a
correct way (see \cite{Wight}). Besides, the time development of
quantum field system cannot be obtained. The only way to overcome
these problems is to construct the $S$-matrix which contains all
elastic and inelastic scattering amplitudes of free particles from
the time $t\to-\infty$ to the time $t\to\infty$. It is important
that the $S$-matrix is a unitary operator on the Fock space. It
means that the bound states like positronium, which is a unstable
particle, cannot belong to any Fock space in principle. In addition,
our computing abilities are restricted to the perturbation theory.

Nevertheless, we believe that the $S$-matrix amplitudes should
contain some correct information on possible bound states. The
simplest way to realize this idea is to postulate that a bound state
is a simple pole of an  elastic scattering amplitude of constituent
particles with appropriate quantum numbers. It means, that the
amplitudes out of mass shell and out of perturbation approach should
be calculated. Standard methods to go out of perturbation
calculations are reduced to sum appropriate classes of Feynman
diagrams and this summation can be formulated in a form of integral
equations. The best known approaches are the Bethe-Salpeter and
Schwinger-Dyson equations. There is numerous literature devoted
these equations (see, for example,
\cite{Wick,Nak,Ef-bos,Ef-ferm,Dor,Ef-scal,Ef-gauge}). The important
difference comparable with the nonrelativistic case is that bound
states in these equations are created by particles which are out of
mass shell so that the role of time becomes important.

One remark on these equations. We know that the perturbation series
are asymptotic series so that the problem is how to sum them? The
exact amplitudes should have some singularity at the point
$\alpha=0$ in QED (see \cite{Dyson}). What is a precise character of
this singularity is not known up to now. Standard perturbation
expansions are connected with Feynman diagrams. Usual  methods are
reduced to summation of an appropriate class of Feynman diagrams.
Result of a summation of any definite class of Feynman diagrams is a
kind of geometrical progression, i.e. it is an analytic function at
the point $\alpha=0$. However, it should be stressed that the
generally accepted point of view - non-perturbed behavior is a sum
of a definite class of Feynman diagrams - is not true.

One of probably successful proposals to calculate the relativistic
corrections to bound state problem is the so-called non-relativistic
QED (NRQED) (see \cite{Lepage}). The basic idea is that the QM is
correct, only non-relativistic momenta are responsible for bound
state properties. In other words, the Hamiltonian should not depend
on time and the problem is to find somehow relativistically small
corrections to the non-relativistic Coulomb potential. The basic
idea is that for small coupling constants the Born approximation is
a good approximation which is directly defined by the Fourier
transform of the potential. The aim is to extract from the
relativistic $S$-matrix some relativistic corrections to
non-relativistic Hamiltonian. The hypothesis is that the scattering
amplitudes in the non-relativistic Schr\"{o}dinger theory and the
relativistic $S$-matrix theory should coincide in the low energy
limit. The procedure is to write down the non-relativistic
Lagrangian with a set of all possible terms, and coefficients in
front of them are calculated by identification with appropriate
amplitudes of relativistic $S$-matrix. This prescription allows one
to remove effectively  time out of the relativistic equations, in
other words, to place all intermediate particles on their mass
shell. It seems NRQED is supported by experimental data.

Another quantum field idea is that a bound state is defined by an
asymptotic behavior of the vacuum mean value of the corresponding
relativistic currents (see, for example, \cite{Alfaro}) with desired
quantum numbers:
\begin{eqnarray}\label{current}
&&\left\langle0|{\bf J}(x){\bf J}(0)|0\right\rangle=\sum\limits_n
\left\langle0|{\bf J}(x)|n\right\rangle\left\langle n|{\bf
J}(0)|0\right\rangle=\sum\limits_n e^{-E_n|x|} |\left\langle0|{\bf
J}(0)|n\right\rangle|^2\nonumber\\
&&\sim e^{-M_{min}|x|} |\left\langle0|{\bf J}(0)|min\right\rangle|^2
~~~~{\rm for}~~~|x|\to\infty.
\end{eqnarray}
This formula gives a possibility to calculate the mass of the lowest
bound state $|min\rangle$ if $M_{min}<2m$. Essentially, the space of
states $\{|n\rangle\}$ is supposed to contain possible bound states
although we saw that the Fock space cannot contain unstable bound
states. These vacuum mean values (\ref{current}) can be represented
in closed forms by functional methods. The functional methods permit
one to get formally the exact representations for Green functions
which are not connected directly with Feynman diagrams, so that it
is possible to go out of standard perturbation expansions using
asymptotic methods. Development of functional methods permits one to
get the exact character of non-analyticity at the point $\alpha=0$
and to clarify the role of "time" in bound state formation. Exactly
this approach will be used in this paper.

The practically unique experimental object to investigate the bound
state problems is the positronium which is the result of pure QED
interaction. On the one hand, the positronium is not a stable state.
It cannot belong to the asymptotic Fock space. Nevertheless, it
exists. The binding energy of positronium itself is not measured
with great accuracy but the mass difference of two possible states,
ortho-positronium $(1^3S_1)$ and para-positronium $(1^1S_0)$, is
known with very large accuracy
\begin{eqnarray}\label{bindingOP}
&&\Delta \epsilon=\epsilon_{ortho}-\epsilon_{para}
=203.38910~GHz=8.4115\cdot10^{-4}~eV\\
&&=0.580487~\alpha^4m_e={7\over12}\alpha^4m_e\cdot0.99512...\nonumber
\end{eqnarray}
The main contribution  can be explained by the non-relativistic
Breit potential approach (see, for example,
\cite{Ahiezer,Ber,Sap,Eid}) taking into account scattering and
annihilation channels
$$\Delta \epsilon=\epsilon_{ortho}-\epsilon_{para}={7\over12}~\alpha^4m_e,~~~~~~
{7\over12}=\left({1\over3}\right)_{scatt}+\left({1\over4}\right)_{annih}.$$

If we apply the relativistic current formula (\ref{current}) to the
positronium problem, we can write
$$\left\langle0|{\bf J}(x){\bf J}(0)|0\right\rangle=
\sum\limits_{particles} e^{-iE_n|x|} |\left\langle0|{\bf
J}(0)|n\right\rangle|^2+\sum\limits_{photons} e^{-iE_n|x|}
|\left\langle0|{\bf J}(0)|n\right\rangle|^2$$
where
$$\sum\limits_{particles} e^{-iE_n|x|} |\left\langle0|{\bf
J}(0)|n\right\rangle|^2\sim e^{-M_{lowest}|x|}$$ and annihilation
channel looks like
$$\sum\limits_{photons} e^{-iE_n|x|} |\left\langle0|{\bf
J}(0)|n\right\rangle|^2\sim{1\over|x|^2}$$ It means that the
annihilation channel does not take part in the bound state formation
in contradiction with the non-relativistic potential approach.

Another point: we want to understand what is the role of TIME in
formation of bound states.

In this paper we apply functional methods to calculate the
asymptotic behavior of  vacuum mean value (\ref{current}) of
relativistic currents for positronium and clarify  the role of time
in the formation of bound states.

\section{Lagrangian and bound states}

All our calculations will be performed in the Euclidean space.  The
Lagrangian  of the electron field $\psi$ and the electromagnetic
photon field $A_\mu$ looks like
\begin{eqnarray}\label{L}
&& L=-{1\over4}F_{\mu\nu}^2(x)+
(\overline{\psi}(x)[i(\hat{p}+e\hat{A}(x))-m]\psi(x)),\\
&& F_{\mu\nu}(x)=\partial_\mu A_\nu(x)-\partial_\nu
A_\mu(x).\nonumber
\end{eqnarray}
The electron propagator has the standard form
\begin{eqnarray}\label{S}
&& S(x-x')=\left\langle{\rm T}\left[
\psi(x)\overline{\psi}(x')\right]\right\rangle =
\int{dp\over(2\pi)^4}{e^{ip(x-x')}\over m-i\hat{p}}
\end{eqnarray}
The propagator of the photon vector field is
\begin{eqnarray}
\label{D} && D_{\mu\nu}(x-y)=\langle
A_\mu(x)A_\nu(y)\rangle=\delta_{\mu\nu}D(x-y)+{\partial^2\over\partial
x_\mu\partial
x_\nu}D_d(x-y),\\
&&  D(x)=\int{dk\over(2\pi)^4}\cdot{e^{ikx}\over
k^2}={1\over(2\pi)^2x^2},~~~~~D_d(x)=\int{dk\over(2\pi)^4}\cdot{e^{ikx)}\over
k^2}{d(k^2)\over k^2}.\nonumber
\end{eqnarray}

\subsection{Two-point Green function}

The object of our interest is the gauge invariant two-point Green
function
\begin{eqnarray}
\label{G} {\bf G}_\Gamma(x-y)&=&\int\!\!\int {D\overline{\psi}D\psi
DA\over C}e^{-{1\over2}(A_\mu
D^{-1}_{\mu\nu}A_\nu)+(\overline{\psi}[i(\hat{p}+e\hat{A})-m]\psi)
}\nonumber\\
&\cdot&
(\overline{\psi}(x)\Gamma\psi(x))(\overline{\psi}(y)\Gamma\psi(y))
\end{eqnarray}
Here $\Gamma$ is a Dirac matrix which defines the local vertex with
quantum numbers of the state $J_\Gamma=(\overline{\psi}\Gamma\psi)$.
We have for para-positronium $\Gamma=i\gamma_5$ and for
ortho-positronium $\Gamma=\gamma_\mu$.

After integration over the electron fields $\psi$ and
$\overline{\psi}$ we get
\begin{eqnarray}
\label{G0} && {\bf G}_\Gamma(x-y)={\bf B}_\Gamma(x-y)+{\bf
H}_\Gamma(x-y),
\end{eqnarray}
where
\begin{eqnarray}\label{B}
&&{\bf B}_\Gamma(x-y)=\int {DA\over C} e^{-{1\over2} (A_\mu
D^{-1}_{\mu\nu}A_\nu)+{\rm T}[A]}\cdot {\rm Tr}[\Gamma
S(x,y|A)\Gamma S(y,x|A)],
\end{eqnarray}
and
\begin{eqnarray*}
&&{\bf H}_\Gamma(x-y)=\int {DA\over C} e^{-{1\over2} (A_\mu
D^{-1}_{\mu\nu}A_\nu)+{\rm T}[A]}\cdot {\rm Tr}[\Gamma
S(x,x|A)]\cdot{\rm Tr}[\Gamma S(y,y|A)].
\end{eqnarray*}

Here $S(x,y|A)$ is the electron propagator in the external field
$A_\mu$:
\begin{eqnarray}
\label{Pr-el} && S(x,y|A)={1\over
i(\hat{p}+e\hat{A}(x))-m}\delta(x-y)
\end{eqnarray}

The functional
\begin{eqnarray*}
&&{\rm T}[A]={\rm Tr}\ln{i(\hat{p}+e\hat{A})-m\over i\hat{p}-m}={\rm
Tr}\ln\left[1+ie\hat{A}{1\over
i\hat{p}-m}\right]\\
&&={e^2\over2}{\rm Tr}\left[\hat{A}{1\over
i\hat{p}-m}\hat{A}{1\over i\hat{p}-m}\right]+O(e^4A^4)\\
&&={e^2\over2}\int\!\!\!\int dxdy~A_\mu(x)\Pi_{\mu\nu}(x-y)A_\nu(y)+O(e^4A^4),\\
&&S_0(x-y)={1\over i\hat{p}-m}\delta(x-y)=\int
{dp\over(2\pi)^4}{e^{-ip(x-y)}\over i\hat{p}-m},\\
&&\Pi_{\mu\nu}(x-y)={\rm Tr}\left[\gamma_\mu S_0(x-y)\gamma_\nu
S_0(y-x)\right],
\end{eqnarray*}
describes radiation corrections to the photon propagator and to the
photon-photon interaction. In this paper, we neglect this term
because it does not contain spin-spin interaction and, therefore,
does not contribute to ortho- para- positronium mass difference in
the lowest corrections.

\begin{figure}[ht]
\begin{center}
\epsfig{figure=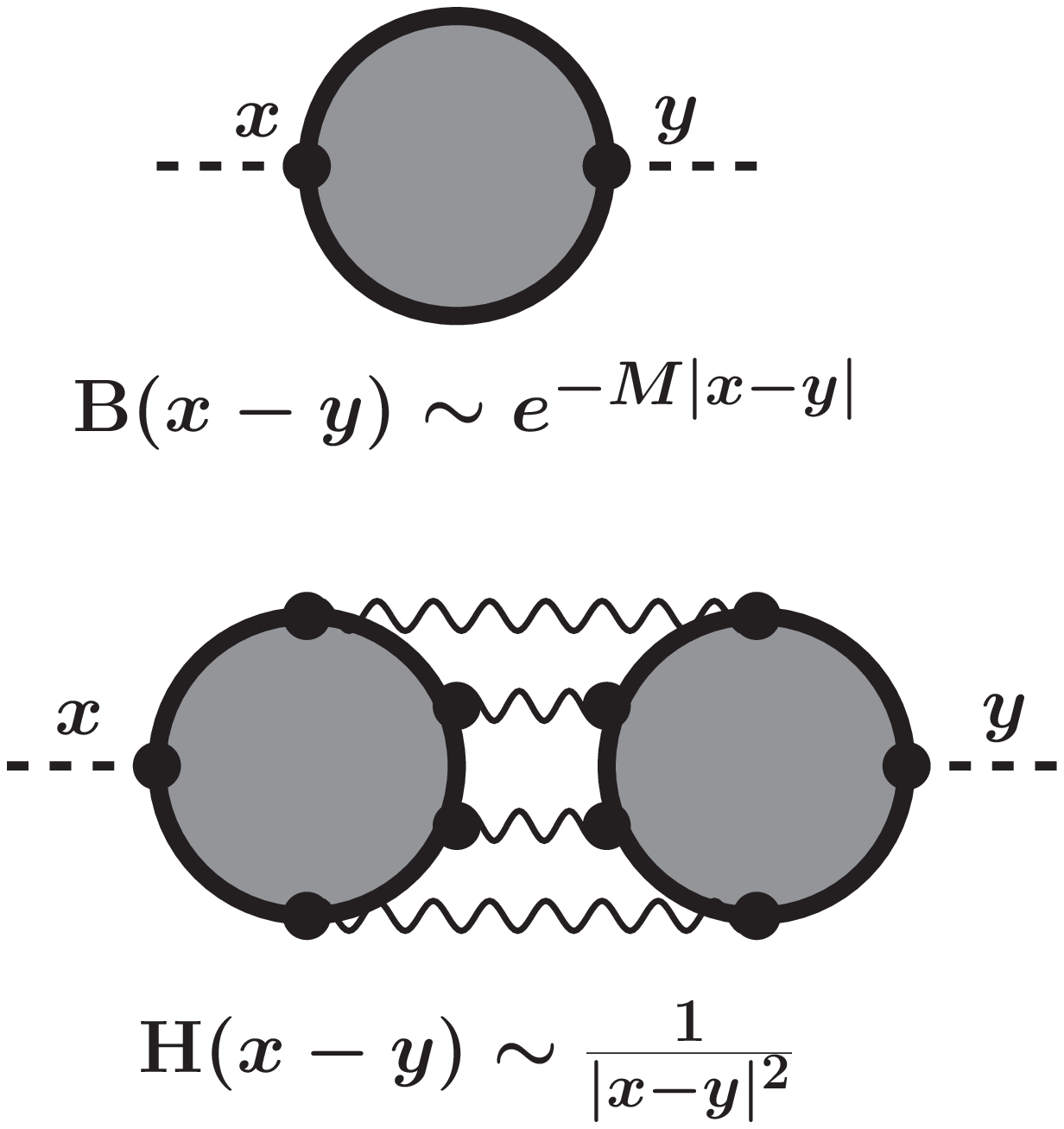,width=8cm}
\end{center}
\caption{Terms  ${\bf B}$ and ${\bf H}$}
\end{figure}

The loop ${\bf B}_\Gamma$ contains all possible
$(\overline{\psi}\Gamma\psi)$-bound states. If the mass of the
lowest state $M_\Gamma<2m$, then the asymptotic behavior of this
loop for large $|x-y|$ looks like
\begin{eqnarray}
\label{PiM} && {\bf B}_\Gamma(x-y)\sim e^{-M_\Gamma|x-y|}
\end{eqnarray}
where $M_\Gamma$ is the mass of the lowest state in the current
$(\overline{\psi}\Gamma\psi)$, i.e. the mass of a possible bound
state. This mass can be calculated by the formula
\begin{eqnarray}
\label{M0} && M_\Gamma=-\lim\limits_{|x|\to\infty}{1\over|x|}\ln{\bf
B}_\Gamma(x)= 2m-\epsilon_\Gamma.
\end{eqnarray}
Here $\epsilon_\Gamma$ defines the binding energy of the lowest
bound state. Our aim is to calculate the functional integral
(\ref{B}) in the limit $|x-y|\to\infty$ and to find $M_\Gamma$,
according to (\ref{M0}).

The loop ${\bf H}$ describes so the called annihilation channel and
contains long-range contributions of photons:
\begin{eqnarray*}
&& {\bf H}_\Gamma(x-y)\sim {1\over|x-y|^2}
\end{eqnarray*}
This term does not contain any bound state.

Graphic representations of the loops ${\bf B}_\Gamma$ and ${\bf
H}_\Gamma$ are shown on Fig.1.

\section{The electron propagator}

The propagator of the electron fermion field satisfies the equation
\begin{eqnarray}
\label{Pr}&& [i(\hat{p}+e\hat{A}(x))-m]S(x,y|A)=\delta(x-y),
\end{eqnarray}
For the gauge transformation
\begin{eqnarray*}
&& A_\mu(x)~\longrightarrow~A_\mu(x)+\partial_\mu f(x)
\end{eqnarray*}
it is transformed as
\begin{eqnarray*}
&& S(x,y|A+\partial f)= e^{ief(x)}S(x,y|A)e^{-ief(y)},
\end{eqnarray*}
so that the loop (\ref{B}) is gauge invariant.

The solution of the equation (\ref{Pr}) can be represented by the
functional integral (see, for example, \cite{Diney}):
\begin{eqnarray}
\label{PrS}
&& S(x,y|A)={1\over i(\hat{p}+e\hat{A}(x))-m}\delta(x-y)\nonumber\\
&& =[i(\hat{p}_x+e\hat{A}(x))+m]\cdot{1\over(p+eA(x))^2+
{e\over2}\sigma_{\mu\nu}F_{\mu\nu}(x)+m^2}\delta(x-y),
\nonumber\\
&&=[i(\hat{p}_x+e\hat{A}(x))+m]\int\limits_0^\infty
{ds\over8\pi^2s^2}e^{- {1\over2}\left[m^2 s +{(x-y)^2\over s}\right]}\nonumber\\
&& \cdot \int{D\eta\over C}e^{-\int\limits_0^s
dt{\dot{\eta}^2(t)\over2}+ie \int\limits_0^s
dt\dot{z}_\mu(t)A_\mu(z(t))} {\rm T}_t\left\{ e^{{e\over4}
\int\limits_0^s dt\sigma_{\mu\nu}(t)F_{\mu\nu}(z(t))}\right\},
\end{eqnarray}
$$z(t)=x{t\over s}+y\left(1-{t\over s}\right)+\eta(t).$$
The boundary conditions are $\eta(0)=\eta(\alpha)=0$ and the
normalization is
$$\int {D\eta\over C}\exp\left\{-\int\limits_0^s
dt{\dot{\eta}^2(t)\over2} \right\}=1.$$ The symbol ${\rm T}_t$ means
the time-ordering of the matrix $\sigma_{\mu\nu}(\tau)$ to the time
variable $t$.

The representation (\ref{PrS}) is quite close to functional
representation of the propagator for a scalar charged particle (see
\cite{Ef-scal}). The main functional structure is the same.

The representation (\ref{PrS}) is obviously gauge covariant because
$$\delta\int\limits_0^s dt~\dot{z}_\mu(t)A_\mu(z(t))=
\int\limits_0^s dt~\dot{z}_\mu(t){\partial\over \partial
z_\mu}f(z(t))$$
$$=\int\limits_0^s dt{d\over dt}f(z(t))=f(z(s))-f(z(0))=f(x)-f(y).$$

As it was said above, our aim is to calculate the functional
integral (\ref{B}) in the limit $|x|\to\infty$ (we put $y=0$). We
want to calculate the main contributions to the binding energy
assuming the coupling constant $\alpha$ to be small. In this case,
for large $x\to\infty$ and small $\alpha$ the saddle-point  in the
integral over $s$  is realized for $s={X\over m}$. Putting
$$x=({\bf x},x_4)\Rightarrow({\bf 0},x_4),~~~~~~~~
\sqrt{x^2}\Rightarrow x_4=X>0,~~~~~t={X\over m}\tau,$$ one can get
for $X\to\infty$
\begin{eqnarray}
\label{PropAs} && S(x,0|A)
\Rightarrow{{\rm const}\over X^{{1\over2}}}(1+\gamma_0)e^{- mX}\cdot{\cal S}(x)\nonumber\\
&& S(0,x|A)
\Rightarrow{{\rm const}\over X^{{1\over2}}}(1-\gamma_0)e^{- mX}\cdot{\cal S}(x)\\
&&{\cal S}(x)=\int{D\eta\over C}e^{-\int\limits_0^X
d\tau~{m\dot{\eta}^2(\tau)\over2}+ie \int\limits_0^X
d\tau~\dot{z}_\mu(\tau)A_\mu(z(\tau))}R[z],\nonumber\\
&&R[z]={\rm T}_\tau\left\{ e^{{e\over4m} \int\limits_0^X
d\tau~\sigma_{\mu\nu}(\tau)F_{\mu\nu}(z(\tau))}\right\}\nonumber
\end{eqnarray}
with
\begin{eqnarray}
\label{z}
&&z(\tau)=n\tau+\eta(\tau)=\left\{\begin{array}{l}
{\mbox{\boldmath$\eta$}}(\tau),\\
\tau+\eta_4(\tau).\\
\end{array}\right.
\end{eqnarray}
We shall use this representation in what follows.

\subsection{Mass of the bound state}

The next step is to substitute electron propagators $S(x,0|A)$ and
$S(0,x|A)$ in the form (\ref{PropAs}) into the representation (\ref{B})
for the Green function ${\bf B}_\Gamma(x)$  and then to integrate
over the photon field $A$. We have for large $X\to\infty$
\begin{eqnarray}
\label{B0} && {\bf B}_\Gamma(X)\sim
e^{-2mX}\int\!\!\!\int{D\eta_1D\eta_2\over
C}e^{-{m\over2}\int\limits_0^X d\tau[\dot{\eta_1}^2(\tau)+
\dot{\eta}_2^2(\tau)]}{\cal F}_\Gamma[X,\eta_1,\eta_2],
\end{eqnarray}
with
\begin{eqnarray}\label{FX}
&&{\cal F}_\Gamma[X,\eta_1,\eta_2]\\
&&=\int {DA\over C} e^{-{1\over2} (A_\mu D^{-1}_{\mu\nu}A_\nu)+ie
\int\limits_0^X d\tau\dot{z}^{(1)}_\mu(\tau)A_\mu(z^{(1)}(\tau))+ie
\int\limits_0^X d\tau\dot{z}^{(2)}_\mu(\tau)A_\mu(z^{(2)}(\tau))}\nonumber\\
&&\cdot{1\over4}{\rm Tr}\left[\Gamma~
\left(1+\gamma_0\right)R[z^{(1)}]~\Gamma~\left(1-\gamma_0\right)
R[z^{(2)}]\right].\nonumber
\end{eqnarray}

The mass $M_\Gamma$ of the bound state with quantum number $\Gamma$
is defined by the formula (\ref{M0}).

The integral (\ref{FX}) over the photon field $A$ can be calculated
explicitly. The result of the calculations is shown in Fig.2. We
will not write down this simple long formula. Our aim is to find the
ortho-para mass difference in the lowest approximation of the
functional method. Therefore, we omit all terms connected with
contributions to the electron propagator in the loop and take into
account the dominant terms responsible for positronium formation and
desired ortho-para mass difference (see Fig.3). We get in the lowest
approximation over spin-spin interaction
\begin{figure}[ht]
\begin{center}
\epsfig{figure=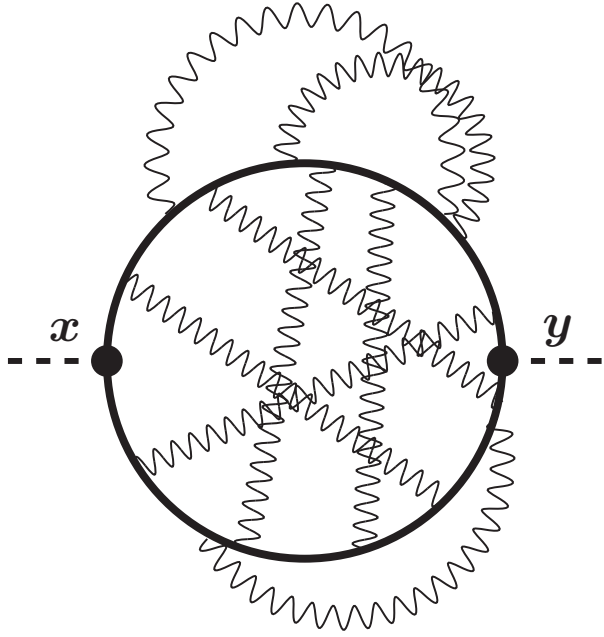,width=8cm}
\end{center}
\caption{All diagrams contributing to (\ref{FX})}
\end{figure}

\begin{figure}[ht]
\begin{center}
\epsfig{figure=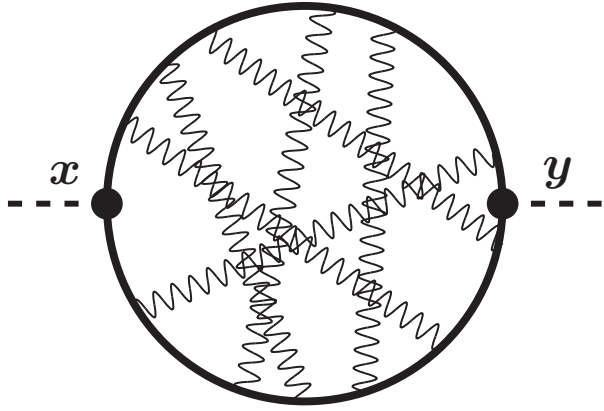,width=8cm}
\end{center}
\caption{Diagrams which are responsible for the bound state}
\end{figure}

\begin{eqnarray*}
&&{\cal F}_\Gamma[X;\eta_1,\eta_2]\\
&&=\int {DA\over C} e^{-{1\over2} (A_\mu D^{-1}_{\mu\nu}A_\nu)}\cdot
e^{ie\int\limits_0^Xd\tau~\dot{z}^{(1)}_\mu(\tau)A_\mu(z^{(1)}(\tau))
+ie\int\limits_0^X d\tau~\dot{z}^{(2)}_\mu(\tau)A_\mu(z^{(2)}(\tau))}\\
&&\cdot{1\over4}{\rm Tr}\Biggl\{\Gamma(1+\gamma_0)\Gamma(1-\gamma_0)\\
&&+{e^2\over16m^2} \int\!\!\!\int\limits_0^Xd\tau_1d\tau_2\cdot
\Gamma(\gamma_0+1)\sigma_{\mu\nu}\Gamma(-\gamma_0+1)\sigma_{\rho\sigma}\cdot
F_{\mu\nu}(z^{(1)}(\tau_1))F_{\rho\sigma}(z^{(2)}(\tau_2)) \Biggr\}\\
&&=e^{W[X;\eta_1,\eta_2]}\cdot\left\{\Sigma_\Gamma^{(0)}+K_\Gamma[s;\eta_1,\eta_2]+O(e^4)\right\}.
\end{eqnarray*}
Here
$$\Sigma_\Gamma^{(0)}={1\over4}{\rm Tr}~\Gamma(1+\gamma_0)\Gamma(1-\gamma_0).$$
The main functional responsible for the bound state formation looks
like
\begin{eqnarray}\label{WX}
&&W[X;\eta_1,\eta_2]=e^2 \int\!\!\!\int\limits_0^Xd\tau_1 d\tau_2~
\dot{z}^{(1)}_\mu(\tau_1)\dot{z}^{(2)}_\nu(\tau_2)D_{\mu\nu}(z^{(1)}(\tau_1)-z^{(2)}(\tau_2))\nonumber\\
&&=e^2 \int\!\!\!\int\limits_0^Xd\tau_1 d\tau_2~
\dot{z}^{(1)}_\mu(\tau_1)\dot{z}^{(2)}_\mu(\tau_2)D(z^{(1)}(\tau_1)-z^{(2)}(\tau_2))\\
&&=e^2\int\!\!\!\int\limits_{0}^Xd\tau_1
d\tau_2~\dot{z}_\mu^{(1)}(\tau_1)\dot{z}_\mu^{(2)}(\tau_2)
\int{dk\over(2\pi)^4}\cdot {e^{ik(z^{(1)}(\tau_1)- z^{(2)}(\tau_2))}\over k^2}.\nonumber
\end{eqnarray}
The functional responsible for the ortho- para- mass difference is
\begin{eqnarray}\label{K}
&&K_\Gamma[X;\eta_1,\eta_2]={e^2\over4m^2}
\int\!\!\!\int\limits_0^Xd\tau_1 d\tau_2\cdot{1\over4}{\rm
Tr}~\Gamma(1+\gamma_0)\sigma_{\mu\rho}\Gamma(1-\gamma_0)\sigma_{\nu\rho}\nonumber\\
&&\cdot{\partial^2\over\partial
z^{(1)}_\mu\partial z^{(2)}_\nu}D(z^{(1)}(\tau_1)-z^{(2)}(\tau_2))\\
&&={e^2\over4m^2} \int\!\!\!\int\limits_0^X d\tau_1
d\tau_2\int{dk\over(2\pi)^4}\cdot e^{ik(z^{(1)}(\tau_1)-
z^{(2)}(\tau_2))}\Sigma_\Gamma(k),\nonumber
\end{eqnarray}
where
\begin{eqnarray}\label{Sigma}
&& \Sigma_\Gamma(k)={k_\mu k_\rho\over k^2}\cdot {1\over4}{\rm
Tr}~\Gamma(1+\gamma_0)\sigma_{\mu\nu}\Gamma(1-\gamma_0)\sigma_{\rho\nu}=
\Sigma_\Gamma^{(0)}\cdot\Delta_\Gamma(k).
\end{eqnarray}

The term $W$ is responsible for positronium bound states. The term
$K_\Gamma$ describes the spin-spin interaction and it is responsible
for the ortho- and para- mass difference. All neglected terms give
the next to $\alpha={e^2\over4\pi}$ perturbation contributions.

Let us introduce the notation
\begin{eqnarray}\label{dsigma}
&& d\sigma_{\eta_1\eta_2}^W={D\eta_1D\eta_2\over
C}e^{-{m\over2}\int\limits_0^Xdt[\dot{\eta_1}^2(t)+
\dot{\eta}_2^2(t)]+W[X;\eta_1,\eta_2]}
\end{eqnarray}
with
\begin{eqnarray}\label{J0X}
&& J_0(X)=\int d\sigma_{\eta_1\eta_2}^W=J e^{\epsilon_0 X}
\end{eqnarray}
then one can write
\begin{eqnarray}\label{BX}
&&{\bf B}_\Gamma(X)=e^{-2mX}\int\!\!\!\int d\sigma_{\eta_1\eta_2}^W
\cdot\left\{\Sigma_\Gamma^{(0)}+K_\Gamma[X;\eta_1,\eta_2]+O(e^4)\right\}\nonumber\\
&&=J~\Sigma_\Gamma^{(0)}e^{-(2m-\epsilon_0-\epsilon_\Gamma) X}
\end{eqnarray}
The binding energy $\epsilon_0$ does not depend on the spin of
electron-positron system and it is defined mainly by the Coulomb
interaction. The binding energy $\epsilon_\Gamma$ depends on the
spin of constituents and defines the ortho-para mass difference. It
looks like
\begin{eqnarray}\label{egamma}
&&\epsilon_\Gamma=\lim\limits_{X\to\infty}{e^2\over4m^2} {1\over
X}\int\!\!\!\int\limits_0^{X} d\tau_1
d\tau_2\int{dk\over(2\pi)^4}\cdot \left\langle
e^{ik(z^{(1)}(\tau_1)-z^{(2)}(\tau_2))}\right\rangle_{\eta_1\eta_2}\Delta_\Gamma(k),\nonumber\\
\end{eqnarray}
where the average is
\begin{eqnarray}\label{brek}
&&\left\langle
e^{ik(z^{(1)}(\tau_1)-z^{(2)}(\tau_2))}\right\rangle_{\eta_1\eta_2}=
{1\over J_0(X)}\int\!\!\!\int d\sigma_{\eta_1\eta_2}^W
e^{ik(z^{(1)}(\tau_1)-z^{(2)}(\tau_2))}
\end{eqnarray}
The functional $W$ in (\ref{dsigma}) contains terms defining the
Coulomb and "time" interactions. The measure
$d\sigma_{\eta_1\eta_2}^W$ contains the space
${\mbox{\boldmath$\eta$}}$ and "time" $\eta_4$ functional variables.
Our aim is to evaluate the contribution of "time" interaction to the
bound state formation. Thus our direct problem is to calculate the
integral (\ref{brek}).

\section{Para- and ortho-positronium mass difference}

In the quantum field theory there is the problems how to define the
bound state. The point is that the relativistic currents of the type
$(\overline{\psi}\Gamma\psi)$ are classified by the relativistic
group in the space ${\bf R}^4$ while the physical bound states are
classified by the non-relativistic group in the space ${\rm R}^3$.
As a result the physical states are described by an appropriate
mixture of  different components of different relativistic currents.
In addition, the angles of mixture are not known a priori. In Table
2 the non-relativistic quantum numbers of different components of
relativistic currents are listed.

One can see that each current $(\overline{\psi}\Gamma\psi)$ with
quantum number $J^P$ is determined by a mixture of two relativistic
currents
\begin{eqnarray}\label{currents}
&& S(0^+):~~~\Gamma_S=I \cos\theta_S+\gamma_0\sin\theta_S,\nonumber\\
&& {\bf
A}(1^+):~~~\Gamma_A=\gamma_5\mbox{\boldmath$\gamma$}\cos\theta_A+
i[\mbox{\boldmath$\gamma$}\times\mbox{\boldmath$\gamma$}]\sin\theta_A,\nonumber\\
&& {\bf V}(1^-):~~~\Gamma_V=\mbox{\boldmath$\gamma$}\cos\theta_V+
i\gamma_0\mbox{\boldmath$\gamma$}\sin\theta_V,\nonumber\\
&&
P(0^-):~~~\Gamma_P=i\gamma_5\cos\theta_P+\gamma_5\gamma_0\sin\theta_P.
\end{eqnarray}
Generally speaking,in order to define the angles
$\theta_S,~\theta_A,~\theta_V,~\theta_P$, some additional
argumentation should be used. However, we shall see that the desired
masses do not depend on these angles at least in the lowest
approximation.

Let us come to formula (\ref{Sigma}). The results for
$\Sigma_\Gamma^{(0)}$ and $\Sigma_\Gamma(k)$ with currents
(\ref{currents}) are listed in Table 3. One can see that the states
$S(0^+)$ and $A(1^+)$ do not exist at all. The masses of bound
states $P(0^-)$ and $V(1^-)$ do not depend on the mixing angles
$\theta_P$ and $\theta_V$.

According to (\ref{egamma}), the desired mass difference is defined
by the formula
\begin{eqnarray}\label{MassDif}
&&\delta M=\epsilon_V-\epsilon_P\nonumber\\
&&={e^2\over4}\lim\limits_{X\to\infty} {1\over X}
\int\!\!\!\int\limits_0^X d\tau_1 d\tau_2\int{dk\over(2\pi)^4}\cdot
\left\langle e^{ik(z^{(1)}(\tau_1)-
z^{(2)}(\tau_2))}\right\rangle_\eta(\Delta_V(k)-\Delta_P(k))\nonumber\\
&&= {8\over3}\cdot{e^2\over4}\lim\limits_{X\to\infty} {1\over
X}\int\!\!\!\int\limits_0^X d\tau_1 d\tau_2\int\!\!\!\int{d{\bf
k}dk_4\over(2\pi)^4}\cdot e^{ik_4(\tau_1-\tau_2)}{{\bf k}^2\over {\bf k}^2+k_4^2}\\
&&\cdot\left\langle
e^{ik_4(\eta_1(\tau_1)-\eta_2(\tau_2))}\right\rangle_{\eta_4}
\cdot\left\langle e^{i{\bf
k}({\mbox{\boldmath$\eta$}}_1(\tau_2)-{\mbox{\boldmath$\eta$}}_2(\tau_2)
)}\right\rangle_{{\mbox{\boldmath$\eta$}}}\nonumber
\end{eqnarray}

\section{The lowest contribution}

The lowest main contribution to positronium bound state is defined
by the integrals (\ref{J0X}) and (\ref{brek}). It is convenient to
extract the $\alpha^2$ dependence  in order to extract the
non-relativistic Coulomb potential (see \cite{Ef-scal}). For this
aim in the representation (\ref{dsigma}) let us introduce the new
variables:
$$ Y=\alpha^2m X,~~\tau={v\over \alpha^2m},~~k_4=\alpha^2m q,~~{\bf k}=\alpha m{\bf q}.$$
$$ {\mbox{\boldmath$\eta$}}(t)={1\over\alpha
m}{\mbox{\boldmath$\xi$}}(v), ~~ \eta(t)={1\over \alpha m}\xi(v).$$
The parameters $X$ and $Y$ are infinitely large quantities.

We get
\begin{eqnarray}
\label{J0Y1} && J_0(Y)=\int\!\!\!\int {D{\mbox{\boldmath$\xi$}}_1
D{\mbox{\boldmath$\xi$}}_2 D\xi_1D\xi_2\over C}
e^{-{1\over2}\int\limits_{0}^{Y}dv
\left[\dot{{\mbox{\boldmath$\xi$}}}_1^2(v)+
\dot{{\mbox{\boldmath$\xi$}}}_2^2(v)+\dot{\xi}_1^2(\tau)+
\dot{\xi}_2^2(v)\right]+
W\left[{\mbox{\boldmath$\xi$}}_1,{\mbox{\boldmath$\xi$}}_2,\xi_1,\xi_2;\alpha\right]},\nonumber\\
\end{eqnarray}
with
\begin{eqnarray}\label{WW}
&&W[{\mbox{\boldmath$\xi$}}_1,{\mbox{\boldmath$\xi$}}_2,\xi_1,\xi_2;\alpha]\nonumber\\
&&=\int\!\!\!\int\limits_{0}^Ydv_1 dv_2
\left[\left(1+\alpha\dot{\xi}_1(v_1)\right)\left(1+\alpha\dot{\xi}_2(v_2)\right)+
\alpha^2\dot{{\mbox{\boldmath$\xi$}}}_1(v_1)\dot{{\mbox{\boldmath$\xi$}}}_2(v_2)\right]\nonumber\\
&&\cdot\int\!\!\!\int {d{\bf q}dq\over4\pi^3}
{e^{iq(v_1-v_2)+\alpha(\xi_1(v_1)-\xi_2(v_2)) +i{\bf
q}({\mbox{\boldmath$\xi$}}_1(v_1) -{\mbox{\boldmath$\xi$}}_2(v_2))}
\over {\bf q}^2+\alpha^2q^2}
\end{eqnarray}

It is important to stress that the functional
$W[{\mbox{\boldmath$\xi$}}_1,{\mbox{\boldmath$\xi$}}_2,\xi_1,\xi_2;\alpha]$
is not analytic at the point $\alpha=0$, so that the relativistic
corrections cannot be obtained by a regular method. Calculation of
the integral (\ref{J0Y1}) is not a simple problem. Therefore, we
restrict ourselves to calculation of the next relativistic
correction to $\alpha$, so that we neglect terms with $\alpha$ in
square brackets in (\ref{WW}) and  introduce the variables
\begin{eqnarray*}
&&{\mbox{\boldmath$\xi$}}_1(\tau)={\bf
R}(\tau)+{1\over2}{\mbox{\boldmath$\rho$}}(\tau),~~~~~{\mbox{\boldmath$\xi$}}_2(\tau)={\bf
R}(\tau)-{1\over2}{\mbox{\boldmath$\rho$}}(\tau),\\
&&\xi_1(\tau)=R(\tau)+{1\over2}\rho(\tau),~~~~~\xi_2(\tau)=R(\tau)-{1\over2}\rho(\tau),
\end{eqnarray*}
The variables ${\bf R}$, ${\mbox{\boldmath$\rho$}}$ and $R$, $\rho$
describe  the center of mass and relative coordinates in
configuration and "time"$~$ spaces, respectively. We shall see that
the "time"$~$ variables give the important contribution to the
desired mass correction. We get
\begin{eqnarray}\label{J0Y2}
&& J_0(Y)=\int\!\!\!\int {D{\mbox{\boldmath$\rho$}} D{\bf R} D\rho
DR\over C} e^{-\int\limits_{0}^{Y}dv
\left[{1\over4}\dot{{\mbox{\boldmath$\rho$}}}^2(v)+ \dot{{\bf
R}}^2(v)+{1\over4}\dot{\rho}^2(v)+ \dot{R}^2(v)\right]+{\bf
W}\left[{\bf
R},{\mbox{\boldmath$\rho$}},R,\rho;\alpha\right]}\nonumber\\
\end{eqnarray}
where
\begin{eqnarray}\label{WWW}
&&{\bf W}[{\bf R},{\mbox{\boldmath$\rho$}},R,\rho;\alpha]
=\int\!\!\!\int\limits_{0}^{Y} dv_1 dv_2\\
&&\cdot\int\!\!\!\int{d{\bf q}dq\over4\pi^3}{e^{iq\left(v_1-v_2+
\alpha\left(R(v_1)-R(v_2)+{1\over2}(\rho(v_1)+\rho(v_2)\right)\right)+
i{\bf q}\left({\bf R}(v_1)-{\bf
R}(v_2)+{1\over2}({\mbox{\boldmath$\rho$}}(v_1)+{\mbox{\boldmath$\rho$}}(v_2)\right)}\over
{\bf q}^2+\alpha^2q^2}\nonumber
\end{eqnarray}

The variables  ${\bf R}$ and $R$ are connected with the continuous
spectrum so that in the lowest perturbation order over the
interaction functional $W[{\bf
R},{\mbox{\boldmath$\rho$}},R,\rho;\alpha]$ we can integrate over
variables ${\bf R}$ and $R$. We get for large $X$:
\begin{eqnarray}\label{J0Y3}
&& J_0(Y)=\int{D\rho D{\mbox{\boldmath$\rho$}}\over
C}e^{-\int\limits_{0}^{Y}dv\left[{1\over4}\dot{\rho}^2(v)+
{1\over4}\dot{{\mbox{\boldmath$\rho$}}}^2(v)\right]+
W\left[{\mbox{\boldmath$\rho$}},\rho;\alpha\right]}.
\end{eqnarray}
Here
\begin{eqnarray*}\label{We}
&& W\left[{\mbox{\boldmath$\rho$}},\rho;\alpha\right]=
\int\!\!\!\int {D{\bf R}DR\over C} e^{-\int\limits_{0}^{Y}dv
\left[\dot{{\bf R}}^2(v)+ \dot{R}^2(v)\right]} W\left[{\bf
R},{\mbox{\boldmath$\rho$}},R,\rho;\alpha\right]\\
&&=\int\!\!\!\int\limits_{0}^{Y}dv_1dv_2 \int\!\!\!\int{d{\bf
q}dq\over4\pi^3} {e^{iq(v_1-v_2)-{1\over4}[\alpha^2q^2+{\bf
q}^2]|v_1-v_2|}\over {\bf q}^2+\alpha^2q^2}\cdot e^{-{i\alpha
q\over2}(\rho(v_1)+\rho(v_2))- {i{\bf
q}\over2}({\mbox{\boldmath$\rho$}}(v_1)+{\mbox{\boldmath$\rho$}}(v_2))}.
\end{eqnarray*}
The functional $W\left[{\mbox{\boldmath$\rho$}},\rho;\alpha\right]$
is not analytic at the point $\alpha=0$. Nevertheless, one can
extract the lowest terms to $\alpha$.
\begin{eqnarray*}\label{W}
&& W\left[{\mbox{\boldmath$\rho$}},\rho;\alpha\right]\Rightarrow
W_0\left[{\mbox{\boldmath$\rho$}},\rho;\alpha\right]\nonumber\\
&&=\int\!\!\!\int\limits_{0}^{Y}dv_1dv_2 \int\!\!\!\int{d{\bf
q}dq\over4\pi^3{\bf q}^2} e^{iq(v_1-v_2)-{1\over4}{\bf
q}^2|v_1-v_2|}\cdot e^{-{i\alpha q\over2}(\rho(v_1)+\rho(v_2))-
{i{\bf
q}\over2}({\mbox{\boldmath$\rho$}}(v_1)+{\mbox{\boldmath$\rho$}}(v_2))}\\
&&=\int{d{\bf q}\over2\pi^2{\bf
q}^2}\int\!\!\!\int\limits_{0}^{Y}dv_1dv_2
\delta\left(v_1-v_2-{\alpha\over2}(\rho(v_1)+\rho(v_2))\right)
e^{-{1\over4}{\bf q}^2|v_1-v_2|- {i{\bf
q}\over2}({\mbox{\boldmath$\rho$}}(v_1)+{\mbox{\boldmath$\rho$}}(v_2))}\\
&&=\int{d{\bf q}\over2\pi^2{\bf q}^2}\int\limits_{0}^{Y}{dv
\over\left|1-{\alpha\over2}\rho'(v)\right|}
e^{-{\alpha\over4}{\bf q}^2|\rho(v)|- i{\bf q}{\mbox{\boldmath$\rho$}}(v)}\\
&&=\int{d{\bf q}\over2\pi^2}\left[\int\limits_{0}^{Y}{d\tau
\over{\bf q}^2} e^{i{\bf
q}{\mbox{\boldmath$\rho$}}(v)}-{\alpha\over4}\int\limits_{0}^{Y}dv
|\rho(v)|e^{- i{\bf
q}{\mbox{\boldmath$\rho$}}(v)}+O(\alpha^{1+\delta})\right].
\end{eqnarray*}
Taking into account the correlations (see Appendix III) one can get
\begin{eqnarray*}
\label{J10} && \delta({\mbox{\boldmath$\rho$}}(v))=
\left\langle\delta({\mbox{\boldmath$\rho$}}(v))\right\rangle_{\mbox{\boldmath$\rho$}}
+\left[\delta({\mbox{\boldmath$\rho$}}(v))-
\left\langle\delta({\mbox{\boldmath$\rho$}}(v))\right\rangle_{\mbox{\boldmath$\rho$}}\right]\\
&&={1\over8\pi}+\left[\delta({\mbox{\boldmath$\rho$}}(v))-{1\over8\pi}\right]
\end{eqnarray*}
because
\begin{eqnarray*}
\label{J11} &&
\left\langle\delta({\mbox{\boldmath$\rho$}}(v))\right\rangle_{\mbox{\boldmath$\rho$}}
=\Psi^2_{0}(0)={1\over8\pi}
\end{eqnarray*}
Finally, we have
\begin{eqnarray}
\label{W0} && W_0\left[{\mbox{\boldmath$\rho$}},\rho;\alpha\right]=
\int\limits_{0}^{Y}{dv\over|{\mbox{\boldmath$\rho$}}(v)|}
-{\alpha\over8}\int\limits_{0}^{Y}dv |\rho(v)|+O(\alpha^{1+\delta})
\end{eqnarray}

The measure in (\ref{J0Y3}) is reduced to
\begin{eqnarray}
&& d\sigma^W={D\rho D{\mbox{\boldmath$\rho$}}\over
C}e^{-\int\limits_{0}^{Y}dv\left[{1\over4}\dot{\rho}^2(v)+
{1\over4}\dot{{\mbox{\boldmath$\rho$}}}^2(v)\right]+
W\left[{\mbox{\boldmath$\rho$}},\rho;\alpha\right]}\\
&&\to d\sigma=d\sigma_{\mbox{\boldmath$\rho$}}d\sigma_\rho,\nonumber
\end{eqnarray}
Here
\begin{eqnarray}
&& d\sigma_{\mbox{\boldmath$\rho$}}={D{\mbox{\boldmath$\rho$}}\over
C_{\mbox{\boldmath$\rho$}}}e^{-\int\limits_{0}^{Y}dv\left[{1\over4}\dot{{\mbox{\boldmath$\rho$}}}^2(v)
-{1\over|{\mbox{\boldmath$\rho$}}(v)|}\right]},\nonumber\\
&& d\sigma_\rho={D\rho \over C_\rho}e^{-\int\limits_{0}^{Y}dv\left[
{1\over4}\dot{\rho}^2(v)+{\alpha\over8} |\rho(v)|\right]}.\nonumber
\end{eqnarray}

This measure consists of two components - the standard
non-relativistic Coulomb potential term and the "time" term which is
not taken into account in any usual calculations. This term
corresponds to one-dimension linear potential with coupling constant
$\alpha$ and leads to non-analytical behavior of energy on $\alpha$.

One can calculate (see Appendix III)
\begin{eqnarray}
&&
J_{Coulomb}=\int
d\sigma_{\mbox{\boldmath$\rho$}}=\int{D{\mbox{\boldmath$\rho$}}\over
C_{\mbox{\boldmath$\rho$}}}e^{-\int\limits_{0}^{Y}dv\left[{1\over4}\dot{{\mbox{\boldmath$\rho$}}}^2(v)
-{1\over|{\mbox{\boldmath$\rho$}}(v)|}\right]}\nonumber\\
&&= \sum\limits_{n\ell}(2\ell+1) e^{-X{\alpha^2\over
n^2m}}|\Psi_{n\ell}(0)|^2,\nonumber
\end{eqnarray}
\begin{eqnarray}
&& J_{time}=\int d\sigma_\rho=\int{D\rho \over
C_\rho}e^{-\int\limits_{0}^{Y}dv\left[
{1\over4}\dot{\rho}^2(v)+{\alpha\over8} |\rho(v)|\right]}=
\sum\limits_\kappa
e^{-X\alpha^{2+{2\over3}}\epsilon_\kappa}m|\Phi_\kappa(0)|^2\nonumber\\
\end{eqnarray}
Thus the general spectrum of the Coulomb and "time" potentials is
\begin{eqnarray}\label{Eepsilon}
&& E_{n\kappa}=\left[-{\alpha^2\over n^2}+
\alpha^{2+{2\over3}}\epsilon_\kappa\right]m
\end{eqnarray}

As a result we have the "time excitations", or abnormal states,
connected with the fourth component of 4-dimensional space. These
states appear in solutions of the Bethe-Salpeter equation. Up to now
it is not known exactly these states does or does not exist in
reality. It is the second reason why relativistic QED does not
describe correctly the real bound states.

\section{Mass difference}

The desired mass difference is defined by the formula
(\ref{MassDif}) which can be represented as
\begin{eqnarray*}
&&\delta M={1\over3}\cdot\alpha^4m\cdot \Delta(\alpha),
\end{eqnarray*}
\begin{eqnarray*}
&&\Delta(\alpha)=\lim\limits_{Y\to\infty} {1\over
Y}\int\!\!\!\int\limits_0^Y dv_1 dv_2\int\!\!\!\int {d{\bf
q}dq\over2\pi^3}\cdot{{\bf
q}^2~e^{iq(v_1-v_2)-{1\over4}[\alpha^2q^2+{\bf
q}^2]|v_1-v_2|}\over {\bf q}^2+\alpha^2 q^2}\\
&&\cdot\left\langle e^{- {i{\bf
q}\over2}({\mbox{\boldmath$\rho$}}(v_1)+{\mbox{\boldmath$\rho$}}(v_2))}
\right\rangle_{{\mbox{\boldmath$\rho$}}}\left\langle e^{-{i\alpha
q\over2}(\rho(v_1)+\rho(v_2))} \right\rangle_{\rho}
\end{eqnarray*}

The averaging over the fields $\rho$ and ${\mbox{\boldmath$\rho$}}$
gives (see Appendix III)
\begin{eqnarray*}
&&\left\langle e^{i{{\bf k}\over2}
[{\mbox{\boldmath$\rho$}}(\tau_1)+{\mbox{\boldmath$\rho$}}(\tau_2)]}
\right\rangle_{{\mbox{\boldmath$\rho$}}}
\Rightarrow\sum\limits_{n\ell}e^{-|\tau_1-\tau_2|(E_{n}-E_{00})}
(-1)^\ell(2\ell+1){\bf C}^2_{n\ell}\left({k\over2}\right).\\
&&\left\langle
e^{-i{q\over2}(\rho(\tau_1)+\rho(\tau_2))}\right\rangle_\rho
\Rightarrow\sum\limits_{\kappa}e^{-|\tau_1-\tau_2|(E_{\kappa}-E_{0})}(-1)^\kappa\left|{\cal
A}_{\kappa}\left(\alpha^{{2\over3}}q\right)\right|^2
\end{eqnarray*}

Finally we have
\begin{eqnarray}\label{Delta}
&&\Delta(\alpha) ={1\over\pi^3}\int\limits_0^\infty
d\tau\int\!\!\!\int d{\bf k}dq\cdot{{\bf k}^2\over {\bf
k}^2+\alpha^2 q^2}\cdot e^{iq\tau-{1\over4}[\alpha^2q^2+{\bf
k}^2]|\tau|}\\
&&\cdot\sum\limits_{n\ell}e^{-{\tau\over4}\left(1-{1\over
n^2}\right)}(-1)^\ell(2\ell+1){\bf
C}^2_{n\ell}\left({k\over2}\right) \sum\limits_{\kappa}(-1)^\kappa
e^{-{\tau\alpha^{{2\over3}}\over4}(\epsilon_{\kappa}-\epsilon_{0})}\left|{\cal
A}_{\kappa}\left(\alpha^{{2\over3}}q\right)\right|^2\nonumber\\
&&=\sum\limits_{n=1}^\infty\sum\limits_{\ell=0}^{n-1}
\sum\limits_{\kappa=0}^{?}(-1)^{\ell+\kappa}\Delta_{n\ell\kappa}(\alpha)\nonumber
\end{eqnarray}
After integration over $\tau$ and angles one can get
\begin{eqnarray}
\Delta_{n\ell\kappa}(\alpha)
&=&{32\over\pi^2}\int\!\!\!\int\limits_0^\infty {dkdq~k^4\over
k^2+\alpha^2q^2}\cdot {k^2+\alpha^2q^2+1-{1\over
n^2}+\alpha^{{2\over3}}(\epsilon_\kappa-\epsilon_0)
\over\left(k^2+\alpha^2q^2+1-{1\over
n^2}+\alpha^{{2\over3}}(\epsilon_\kappa-\epsilon_0)\right)^2+16q^2}\nonumber\\
&\cdot&(2\ell+1){\bf C}^2_{nl}\left({k\over2}\right)\left|{\cal
A}_{\kappa}\left(\alpha^{{2\over3}}q\right)\right|^2
\end{eqnarray}
The numerical results are shown in Table 4. For the function
$\Delta(\alpha)$ we get
\begin{eqnarray*}
&&
\Delta_0(0)=\sum\limits_{n=1}^3\sum\limits_{\ell=0}^{n-1}(-1)^\ell\Delta_{n\ell0}(0)=1.0,\\
&&
\Delta_0(\alpha)=\sum\limits_{n=1}^3\sum\limits_{\ell=0}^{n-1}(-1)^\ell\Delta_{n\ell0}(\alpha)=0.9641,
\end{eqnarray*}

\begin{eqnarray*}
&& \Delta(\alpha)=\sum\limits_{n=1}^3\sum\limits_{\ell=0}^{n-1}
\sum\limits_{\kappa=0}^2(-1)^{\ell+\kappa}\Delta_{n\ell\kappa}(\alpha)=0.952754.
\end{eqnarray*}

Obviously, this result is in contradiction with the existing
experimental number $(\Delta=0.99512...)$.

\section{Breit potential approach}

\begin{figure}[ht]
\begin{center}
\epsfig{figure=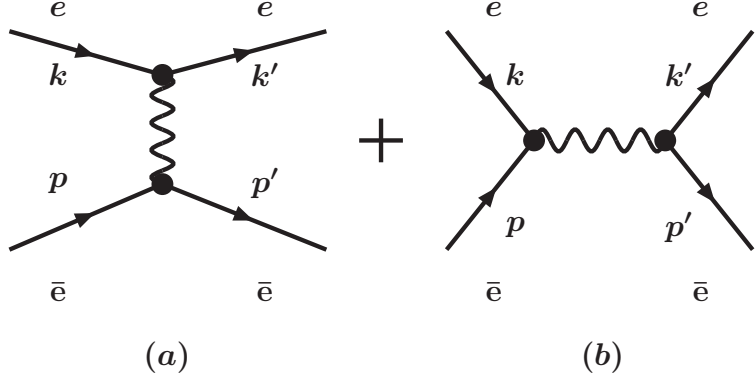,width=10cm}
\end{center}
\caption{Diagrams defining the Breit potential: (a) - scattering and
(b) - annihilation channels.}
\end{figure}

One of the attempts to describe the bound state problem is the Breit
potential approach (see,  for example, \cite{Ahiezer}). Let us
consider the elastic electron-positron scattering
\begin{eqnarray*}
&& e_{p}+\bar{e}_{k}~\Longrightarrow~e_{p'}+\bar{e}_{k'}
\end{eqnarray*}
The scattering amplitude in the lowest order of relativistic
$S$-matrix theory is described by the Feynman diagrams shown on
Fig.2 and looks like
\begin{eqnarray}
\label{ee} M&=&-e^2[\bar{u}(p')\gamma_\mu
u(p)]D_{\mu\nu}(p-p')[\bar{u}(-k)\gamma_\nu u(-k')]\\
&&\nonumber\\
&+&e^2[\bar{u}(-k)\gamma_\mu
u(p)]D_{\mu\nu}(p+k)[\bar{u}(p')\gamma_\nu u(-k')]\nonumber
\end{eqnarray}
where the first term $(a)$ is connected with scattering  and the
second one $(b)$ with annihilation  channels. The spinors
$\bar{u}(p')$ and $u(k')$ are the solutions of the Dirac equation.
This amplitude in the non-relativistic limit should coincide with
the non-relativistic Born approximation which defines the effective
electron-positron potential. This potential should be introduced
into the non-relativistic Schr\"{o}dinger equation.

The Dirac spinors $\bar{u}(p')$ and $u(k')$ define  the relativistic
corrections to the non-relativistic potential. It should emphasize
that electrons and positrons are on the mass shell, i.e. they are
real physical particles. Thus the time is removed from the
interaction Hamiltonian.

The part of the Breit potential which is responsible for the ortho-
and para- mass difference looks like
\begin{eqnarray}
\label{Brpot1} U(r)&=&U_{sc}(r)+U_{an}(r)=-{\alpha\over
r}+{7\over12}\cdot{2\alpha\pi\over
m^2}(\mbox{\boldmath$\sigma$}_-\mbox{\boldmath$\sigma$}_+)\delta({\bf
r}).
\end{eqnarray}
where
\begin{eqnarray}
\label{Brpot2} U_{sc}(r)&=&-{\alpha\over
r}+{1\over3}\cdot{2\alpha\pi\over
m^2}(\mbox{\boldmath$\sigma$}_-\mbox{\boldmath$\sigma$}_+)\delta({\bf
r}),\nonumber\\
U_{an}(r)&=&{1\over4}\cdot{2\alpha\pi\over
m^2}(\mbox{\boldmath$\sigma$}_-\mbox{\boldmath$\sigma$}_+)\delta({\bf
r}).\nonumber
\end{eqnarray}
We want to stress that the coefficient ${7\over12}$ is the sum of
the contributions from scattering and annihilation channels
$${7\over12}=\left({1\over3}\right)_{sc}+\left({1\over4}\right)_{un}$$

Taking into account
\begin{eqnarray*}
&&\Psi^2(0)={\alpha^3m^3\over8\pi},\\
&&\langle(\mbox{\boldmath$\sigma$}_-\mbox{\boldmath$\sigma$}_+)\rangle_{ortho}-
\langle(\mbox{\boldmath$\sigma$}_-\mbox{\boldmath$\sigma$}_+)\rangle_{para}=4,
\end{eqnarray*}
one can get for the ortho- and para- mass difference
\begin{eqnarray*}
&&\Delta=\epsilon_{ortho}-\epsilon_{para}={7\over12}\cdot{2\alpha\pi\over
m^2}\cdot{\alpha^3m^3\over8\pi}\cdot4={7\over12}\alpha^4m
\end{eqnarray*}
This result is in  good agreement with the experimental data and,
therefore, supports the point of view that the annihilation channel
plays the essential role in the formation of the positronium.

\section{Conclusion}

In conclusion, one can say that the functional approach is the best
mathematical representation to preserve the gauge invariance. The
developed technique of calculations permits one to get accurate
results in QED where the coupling constant $\alpha$ is small. The
lowest approximation of this functional representation is the pure
non-relativistic Feynman path integral representation of the
non-relativistic Schr\"{o}dinger equation with the Coulomb
potential. One can see that any regular series for next corrections
to $\alpha$ do not exist and these corrections can not be reduced to
some terms to the non-relativistic potential in the Schr\"{o}dinger
picture. In other words,the "nonphysical"$~$ time coordinate is
important and leads to corrections which is not analytic of the
order $\alpha^{{2\over3}}$.

There exists a contradiction in the current algebra formula
(\ref{current}). On one hand, it is supposed that the space of
states $\{|n\rangle\}$ can contain possible bound states. However on
the other hand, in reality it is the Fock space of free electrons
and photons which does not contain any unstable bound states.
Nevertheless calculations of the functional representation for an
appropriate Green function in the limit $t\to\infty$ indicate that a
bound with $M_{bound}<2m$ does exist really. Besides, the current
algebra in QFT excludes influence of the annihilation channel for
the bound state formation.

Our calculations show that the role of time is very important and
give essential contribution into bound state mass. The next
radiation corrections, connected with time excitations, to
electromagnetic mass difference to positronium  are of the order
$\alpha^{{2\over3}}$, i.e. they are to large.

In addition, the "time excitations", or abnormal states arise in QFT
calculations but they are not exist in reality.

The experimental value of ortho- para-positronium mass difference is
described in the framework of the Breit potential picture with
attraction of the annihilation channel. Thus, explanation of
experimental value para- ortho- positronium mass difference requires
to take into account annihilation channel for effective potential.

One can conclude that in the relativistic QED time corrections are
important, but the bound state problem requires the non-relativistic
potential description where the time variable does not play any
essential role.

The conclusion:  the relativistic QED is not suited to describe real
bound states correctly.

{\bf Acknowledgements}

I am grateful to R.N.Faustov for helpful discussions.

\section{Appendix I}

We use the following representation for $\gamma$ matrices:
\begin{eqnarray*}
&&\gamma_0=\left(\begin{array}{cc} 1 & 0\\
0 & -1\\
\end{array}\right),~~~~~
\mbox{\boldmath$\gamma$}=\left(\begin{array}{cc} 0 &\mbox{\boldmath$\sigma$}\\
-\mbox{\boldmath$\sigma$} & 0\\
\end{array}\right),~~~~
\gamma_5=\left(\begin{array}{cc} 0 & 1\\
1 & 0\\
\end{array}\right)
\end{eqnarray*}
and
\begin{eqnarray*}
&&\sigma_{\mu\nu}={1\over2i}(\gamma_\mu\gamma_\nu-\gamma_\nu\gamma_\mu),\\
&&\sigma_{0j}=-i\sigma_j\gamma_5,~~~~~\sigma_{ij}=-\epsilon_{ijk}\sigma_k
\end{eqnarray*}
We should calculate the traces
\begin{eqnarray*}
&& \Sigma^{(0)}_\Gamma= {1\over4}{\rm
Tr}~\Gamma(1+\gamma_0)\Gamma(1-\gamma_0)
\end{eqnarray*}
and
\begin{eqnarray*}
&& \Sigma_\Gamma(k)={k_\mu k_\nu\over k^2}\cdot {1\over4}{\rm
Tr}~\Gamma(1+\gamma_0)\sigma_{\mu\rho}\Gamma(1-\gamma_0)\sigma_{\nu\rho}=
\Sigma_\Gamma^{(0)}\cdot\Delta_\Gamma(k).
\end{eqnarray*}

For the para-positronium $(P)$ with $\Gamma_P=i\gamma_5$ we get
\begin{eqnarray*}
&& \Sigma^{(0)}_P= {1\over4}{\rm
Tr}~i\gamma_5(1+\gamma_0)i\gamma_5(1-\gamma_0)=-2
\end{eqnarray*}
and
\begin{eqnarray*}
&& \Sigma_P(k)={k_\mu k_\nu\over k^2}\cdot {1\over4}{\rm
Tr}~i\gamma_5(1+\gamma_0)\sigma_{\mu\rho}i\gamma_5(1-\gamma_0)\sigma_{\nu\rho}=
{4{\bf k}^2\over {\bf k}^2+k_4^2}
\end{eqnarray*}

For the ortho-positronium $(V)$ with $\Gamma_V=\gamma_j$ we have
\begin{eqnarray*}
&& \Sigma^{(0)}_P= {1\over4}{\rm
Tr}~\gamma_i(1+\gamma_0)\gamma_j(1-\gamma_0)=-2\delta_{ij}
\end{eqnarray*}
and
\begin{eqnarray*}
&& \Sigma_V(k)={k_\mu k_\nu\over k^2}\cdot {1\over4}{\rm
Tr}~\gamma_i(1+\gamma_0)\sigma_{\mu\rho}\gamma_j(1-\gamma_0)\sigma_{\nu\rho}=
-\delta_{ij}{4{\bf k}^2\over 3({\bf k}^2+k_4^2)}
\end{eqnarray*}
The results are collected in Table 2.

\begin{eqnarray*}
&& \Delta_V(k)-\Delta_P(k)={8\over3}\cdot{{\bf k}^2\over {\bf
k}^2+k_4^2}.
\end{eqnarray*}

\section{Appendix II}

Let us consider the contribution of the longitudinal part of the
photon propagator  to the integral (\ref{FX}). We have equality
$$\dot{z}_\mu(\tau)\partial_\mu f(z(\tau))={\partial\over\partial \tau}
f(z(\tau))$$ Then the term with the gauge dependent part
$\partial_\mu\partial_\nu D_d$ looks like
\begin{eqnarray*}
&&W_{ij}^d={e^2\over2}\int\limits_{0}^{s_i}d\tau_1
\int\limits_{0}^{s_j}d\tau_2~\dot{z}_\mu^{(i)}(\tau_1)\dot{z}_\nu^{(j)}(\tau_2)
\partial_\mu\partial_\nu D_d(z^{(i)}(\tau_1)-z^{(j)}(\tau_2))\\
&&={e^2\over2}\int\limits_{0}^{s_i}d\tau_1
\int\limits_{0}^{s_j}d\tau_2~{\partial^2\over\partial\tau_1\partial\tau_2}
D_d(z^{(i)}(\tau_1)-z^{(j)}(\tau_2))\\
&&={e^2\over2}\left[D_d(z^{(i)}(s_i)-z_\mu^{(j)}(s_j))-D_d(z^{(i)}(0)-z^{(j)}(s_j))\right.\\
&&\left.-D_d(z^{(i)}(s_i)-z_\mu^{(j)}(0))+D_d(z^{(i)}(0)-z^{(j)}(0))\right]\\
&&=e^2\left[D_d(0)-D_d(x-y)\right]=e^2\int{dk\over(2\pi)^4}{d(k^2)\over
k^2}{1-e^{ik(x-y)}\over k^2}
\end{eqnarray*}
This term does not contribute to the bound state mass and should be
omitted.

\section{Appendix III}

We consider the integral
\begin{eqnarray}\label{A1}
&&I=\int d\sigma ~e^{\Phi}=
\int\limits_{{\mbox{\boldmath$\rho$}}(0)=0,~{\mbox{\boldmath$\rho$}}(t)=0}
{D{\mbox{\boldmath$\rho$}}\over C} e^{-\int\limits_{0}^td\tau
\left[{1\over4}\dot{{\mbox{\boldmath$\rho$}}}^2(\tau)-U(|{\mbox{\boldmath$\rho$}}(\tau)|)
\right]+\Phi[{\mbox{\boldmath$\rho$}}]}\nonumber\\
\end{eqnarray}
where
\begin{eqnarray*}
&&\Phi[{\mbox{\boldmath$\rho$}}]={1\over2}\int\!\!\!\int\limits_0^td\tau_1d\tau_2~A(|\tau_1-\tau_2|)\cdot
e^{i{{\bf
k}\over2}({\mbox{\boldmath$\rho$}}(\tau_1)+{\mbox{\boldmath$\rho$}}(\tau_2))}
\end{eqnarray*}
The assumption is that the functional $\Phi$ is  small
$||\Phi[{\mbox{\boldmath$\rho$}}]||\ll1$. We have in this case
\begin{eqnarray*}
&&I=\int d\sigma ~e^{\Phi}=N\exp\left\{{1\over N}\int
d\sigma~\Phi+O(\Phi^2)\right\},~~~~~N=\int d\sigma.
\end{eqnarray*}
and the problem is to calculate the integral
\begin{eqnarray}\label{A2}
&&W=\langle\Phi[{\mbox{\boldmath$\rho$}}]\rangle={1\over N}\int
d\sigma~\Phi\\
&&={1\over N}
\int\limits_{{\mbox{\boldmath$\rho$}}(0)=0,~{\mbox{\boldmath$\rho$}}(t)=0}
{D{\mbox{\boldmath$\rho$}}\over C} e^{-\int\limits_{0}^td\tau
\left[{1\over4}\dot{{\mbox{\boldmath$\rho$}}}^2(\tau)-U(|{\mbox{\boldmath$\rho$}}(\tau)|)
\right]}\Phi[{\mbox{\boldmath$\rho$}}].\nonumber
\end{eqnarray}

Let the Hamiltonian
\begin{eqnarray*}
&&H=-\left({\partial\over\partial{\bf x}}\right)^2+U(|{\bf x}|)
\end{eqnarray*}
have the spectrum $\{E_n\}$ with the wave functions $\Psi_n({\bf
x})$
\begin{eqnarray*}
&& H\Psi_n({\bf x})=E_n\Psi_n({\bf x}).
\end{eqnarray*}
The time Green function can be represented in two forms
\begin{eqnarray}
\label{A3} && G_{t-t'}({\bf x},{\bf x}')=e^{-H(t-t')}\delta({\bf
x}-{\bf x}')=\int\limits_{{\mbox{\boldmath$\rho$}}(t')={\bf
x}',{\mbox{\boldmath$\rho$}}(t)={\bf x}}
{D{\mbox{\boldmath$\rho$}}\over C} e^{-\int\limits_{t'}^{t}d\tau
\left[{1\over4}\dot{{\mbox{\boldmath$\rho$}}}^2(\tau)-U(|{\mbox{\boldmath$\rho$}}(\tau)|)
\right]}\nonumber\\
&& =\sum\limits_n\Psi_n({\bf x})e^{-(t-t')E_n}\Psi_n^+({\bf x}').
\end{eqnarray}
The Green function satisfies  the correlation for $t>t''>t'$
\begin{eqnarray*}
&&G_{t-t'}({\bf x},{\bf x}') =\int d{\bf y}~G_{t-t''}({\bf x},{\bf
y})G_{t''-t'}({\bf y},{\bf x}')
\end{eqnarray*}

We have for the function $W$
\begin{eqnarray*}
&&W(t)=\int\limits_0^t
d\tau_1\int\limits_0^{\tau_1}d\tau_2~A(|\tau_1-\tau_2|)H(\tau_1,\tau_2)
\end{eqnarray*}
with
\begin{eqnarray*}
&&H(\tau_1,\tau_2)={1\over N} \int d{\mbox{\boldmath$\rho$}}_1 \int
d{\mbox{\boldmath$\rho$}}_2~
G_{t-\tau_1}(0,{\mbox{\boldmath$\rho$}}_1)e^{i{{\bf
k}\over2}{\mbox{\boldmath$\rho$}}_1}
G_{\tau_1-\tau_2}({\mbox{\boldmath$\rho$}}_1,{\mbox{\boldmath$\rho$}}_2)
e^{i{{\bf k}\over2}{\mbox{\boldmath$\rho$}}_2}G_{\tau_2}({\mbox{\boldmath$\rho$}}_2,0)\\
&&=\sum\limits_{n_1n_2n_3}{1\over N} \int\!\!\!\int
d{\mbox{\boldmath$\rho$}}_1  d{\mbox{\boldmath$\rho$}}_2~
e^{-E_{n_1}(X-\tau_1)}\Psi_{n_1}(0)\Psi^*_{n_1}({\mbox{\boldmath$\rho$}}_1)e^{i{{\bf
k}\over2}{\mbox{\boldmath$\rho$}}_1}\\
&&\cdot e^{-E_{n_2}(\tau_1-\tau_2)}
\Psi_{n_2}({\mbox{\boldmath$\rho$}}_1)\Psi^*_{n_2}({\mbox{\boldmath$\rho$}}_2)e^{i{{\bf
k}\over2}{\mbox{\boldmath$\rho$}}_2}e^{-E_{n_3}\tau_2}
\Psi_{n_3}({\mbox{\boldmath$\rho$}}_2)\Psi^*_{n_3}(0)\\
&&={1\over N}\sum\limits_{n_1n_2n_3}
e^{-E_{n_1}(X-\tau_1)-E_{n_2}(\tau_1-\tau_2)-E_{n_3}\tau_2}~\Psi_{n_1}(0)\\
&&\cdot\left(\int
d{\mbox{\boldmath$\rho$}}_1\Psi^*_{n_1}({\mbox{\boldmath$\rho$}}_1)e^{i{{\bf
k}\over2}{\mbox{\boldmath$\rho$}}_1}\Psi_{n_2}({\mbox{\boldmath$\rho$}}_1)\right)\left(
\int d{\mbox{\boldmath$\rho$}}_2
\Psi^*_{n_2}({\mbox{\boldmath$\rho$}}_2)e^{i{{\bf
k}\over2}{\mbox{\boldmath$\rho$}}_2}
\Psi_{n_3}({\mbox{\boldmath$\rho$}}_2)\right)\Psi^*_{n_3}(0)
\end{eqnarray*}
The function $W(t)$ for $t\to\infty$  behaves as $W(t)\sim tW_0$. It
means that in the above stated sum in this limit only terms with
$n_1=n_3=0$ survive:
\begin{eqnarray*}
&&H(\tau_1,\tau_2)\Rightarrow\sum\limits_{n}
e^{-(E_{n}-E_0)(\tau_1-\tau_2)}\\
&&\cdot \left(\int
d{\mbox{\boldmath$\rho$}}_1\Psi^*_{0}({\mbox{\boldmath$\rho$}}_1)e^{i{{\bf
k}\over2}{\mbox{\boldmath$\rho$}}_1}\Psi_{n}({\mbox{\boldmath$\rho$}}_1)\right)\left(
\int d{\mbox{\boldmath$\rho$}}_2
\Psi^*_{n}({\mbox{\boldmath$\rho$}}_2)e^{i{{\bf
k}\over2}{\mbox{\boldmath$\rho$}}_2}
\Psi_{0}({\mbox{\boldmath$\rho$}}_2)\right)\\
&&=\sum\limits_{n} e^{-(E_{n}-E_0)(\tau_1-\tau_2)}C_{0n}\left({{\bf
k}\over2}\right)C^*_{0n}\left(-{{\bf k}\over2}\right)
\end{eqnarray*}
where
\begin{eqnarray*}
&&C_{0n}\left({{\bf k}\over2}\right)=\int d{\bf x}~\Psi_0({\bf
x})e^{i{{\bf k}\over2}{\bf x}}\Psi_{n}({\bf x})
\end{eqnarray*}

Finally, we have for $t\to\infty$
\begin{eqnarray*}
&&W(t)=\int\limits_0^t
d\tau_1\int\limits_0^{\tau_1}d\tau_2~A(|\tau_1-\tau_2|)
\sum\limits_{n}e^{-(\tau_1-\tau_2)(E_{n}-E_{0})} C^2_{0n}\left({{\bf
k}\over2}\right)=t~W_0\\
&&W_0=\int\limits_0^\infty d\tau~A(\tau)
\sum\limits_{n}e^{-\tau(E_{n}-E_{0})} C_{0n}\left({{\bf k}\over2}\right) C^*_{0n}\left(-{{\bf k}\over2}\right)\\
\end{eqnarray*}

Thus,
\begin{eqnarray*}
&&I=\int d\sigma ~e^{\Phi}=\Psi^2_0(0)\cdot
e^{-t(E_0-W_0+O(\Phi^2))}
\end{eqnarray*}

\subsection{Spherically symmetric potentials}

If the potential is spherically symmetric
$U=U(|{\mbox{\boldmath$\rho$}}|)$ then the spectrum $\{E_{n\ell}\}$
and the eigenfunctions are
\begin{eqnarray*}
&&\Psi_{n\ell m}({\mbox{\boldmath$\rho$}})=R_{n\ell}(\rho)Y_{\ell
m}({\bf n}),~~~~\sum\limits_ m Y_{\ell m}^*({\bf n})Y_{\ell m}({\bf
n})={2\ell+1\over4\pi}
\end{eqnarray*}
with ortho-normal conditions
\begin{eqnarray*}
&&\int d{\mbox{\boldmath$\rho$}}~\Psi^*_{n\ell
m}({\mbox{\boldmath$\rho$}})\Psi_{n'\ell'
m'}({\mbox{\boldmath$\rho$}})=\delta_{nn'}\delta_{\ell\ell'}\delta_{mm'},\\
&&\sum\limits_{n\ell m}\Psi_{n\ell
m}({\mbox{\boldmath$\rho$}})\Psi^*_{n\ell
m}({\mbox{\boldmath$\rho$}}')=\delta({\mbox{\boldmath$\rho$}}-{\mbox{\boldmath$\rho$}}')=
{1\over \rho^2}\delta(\rho-\rho')\delta({\bf n}-{\bf n}')
\end{eqnarray*}
For radial functions we get ($\rho=|{\mbox{\boldmath$\rho$}}|$)
\begin{eqnarray*}
&&\int\limits_0^\infty d\rho\rho^2~R_{n\ell}(\rho)R_{n'\ell}(\rho)=\delta_{nn'},\\
&&\sum\limits_{n=0}^\infty
R_{n\ell}(\rho)R_{n\ell}(\rho')={1\over\rho^2}\delta(\rho-\rho')
\end{eqnarray*}
The form-factors looks like
\begin{eqnarray*}
&&C_{0n}\left({{\bf k}\over2}\right)=C_{000,n\ell m}\left({{\bf
k}\over2}\right)=\int d{\mbox{\boldmath$\rho$}}~
\Psi_{000}(\rho)e^{i{{\bf
k}\over2}{\mbox{\boldmath$\rho$}}}\Psi_{n\ell m}({\mbox{\boldmath$\rho$}})\\
&&=\int\limits_0^\infty d\rho\rho^2~R_{00}(\rho)R_{nl}(\rho)\int
{d{\bf n}\over\sqrt{4\pi}}\cdot e^{i{{\bf kn}\over2}\rho}Y_{lm}({\bf n})\\
&&=\sqrt{4\pi} i^\ell Y_{\ell m}({\bf n}_k){\bf
C}_{n\ell}\left({k\over2}\right),\\
\end{eqnarray*}
with
\begin{eqnarray*}
&& {\bf C}_{n\ell}\left({k\over2}\right)=\int\limits_0^\infty
d\rho\rho^2~R_{00}(\rho)R_{nl}(\rho)j_\ell\left({k\rho\over2}\right)\\
&&j_\ell\left({k\rho\over2}\right)=\sqrt{{\pi\over
k\rho}}J_{\ell+{1\over2}}\left({k\rho\over2}\right)
\end{eqnarray*}

Finally we get
\begin{eqnarray*}
&&W_0=\int\limits_0^\infty d\tau~A(\tau)
\sum\limits_{n\ell}e^{-\tau(E_{n\ell}-E_{0})}(-1)^\ell(2\ell+1) {\bf
C}^2_{n\ell}\left({k\over2}\right)
\end{eqnarray*}

\subsection{The Coulomb potential}

In the representation (\ref{A1}) the Hamiltonian is
\begin{eqnarray*}
&&
H\Psi_n({\mbox{\boldmath$\rho$}})=E_n\Psi_n({\mbox{\boldmath$\rho$}}),~~~~~
H=-\left({\partial\over\partial{\mbox{\boldmath$\rho$}}}\right)^2-{1\over|{\mbox{\boldmath$\rho$}}|},\\
\end{eqnarray*}
The Coulomb wave functions
$$\Psi_{n\ell m}({\mbox{\boldmath$\rho$}})=R_{n\ell}(\rho)Y_{\ell m}({\bf n})$$
are solutions of the non-relativistic Schr\"{o}dinger equation
($E_n<0$)
\begin{eqnarray*}
&& \left[{d^2\over d\rho^2}+{2\over \rho}{d\over
d\rho}-{\ell(\ell+1)\over \rho^2}+ {1\over
\rho}-|E_{n\ell}|\right]R_{n\ell}(\rho)=0
\end{eqnarray*}
or according to \cite{Landau} one can put
$\rho={r\over2\sqrt{|E_{n\ell}|}}=nr$ so that we have
\begin{eqnarray*}
&& \left[{d^2\over dr^2}+{2\over r}{d\over dr}-{\ell(\ell+1)\over
r^2}+ {n\over r}-{1\over4}\right]{\cal R}_{n\ell}(r)=0,
\end{eqnarray*}
where the spectrum is
\begin{eqnarray*}
&& n={1\over\sqrt{-4E_{n\ell}}}~~~{\rm
or}~~~E_{n\ell}=E_n=-{1\over4n^2}
\end{eqnarray*}
with $n=1,2,3,...$ and $\ell=1,...,n-1$.

Solutions are
\begin{eqnarray*}
&& {\cal R}_{n\ell}(\rho)={1\over n^2(2\ell+1)!}
\sqrt{{(n+\ell)!\over2(n-\ell-1)!}} \left({\rho\over n}\right)^\ell
e^{-{\rho\over2n}} F\left(-n+\ell+1,2\ell+2,{\rho\over n}\right)
\end{eqnarray*}

\begin{eqnarray*}
&& {\bf C}_{n\ell}(k)=\int\limits_0^\infty
d\rho\rho^2~R_{00}(\rho)R_{nl}(\rho)j_\ell\left({k\rho\over2}\right)
\end{eqnarray*}
Several particular functions are
\begin{eqnarray*}
&& {\bf C}_{10}(k)={16\over(4 + k^2)^2}\\
&& {\bf C}_{20}(k)={256\sqrt{2}~k^2\over(9 + 4 k^2)^3},~~~~{\bf C}_{21}(k)={128\sqrt{6}~ k\over(9 + k^2)^3}\\
&& {\bf C}_{30}(k)={432\sqrt{3}~ k^2(16 + 27 k^2)\over(16 +9 k^2)^4}\\
&& {\bf C}_{31}(k)={288\sqrt{6}~ k(16 + 27 k^2)\over(16 +
9k^2)^4},~~~~{\bf C}_{32}(k)=\sqrt{{6\over5}}\cdot{6912~ k^2\over(16
+9k^2)^4}
\end{eqnarray*}

\subsection{"Time" potential}

We have the integral
\begin{eqnarray}
\label{J12} &&
J_X(\alpha)={e^{{X\over4}}\over8\pi}I_X(\alpha),\nonumber\\
&&I_X(\alpha)=\int{D\rho\over
C}e^{-{1\over4}\int\limits_{0}^{X}d\tau\dot{\rho}^2(\tau)+U[\rho,\alpha]},\\
&&U[\rho,\alpha]=W\left[\rho;\alpha\right]-W\left[\rho;0\right].\nonumber
\end{eqnarray}
In the paper \cite{} it is shown that for small $\alpha$
\begin{eqnarray}
\label{U}
&&U[\rho,\alpha]=W\left[\rho;\alpha\right]-W\left[\rho;0\right]=
-{\alpha\over8}\int\limits_0^X d\tau|\rho(\tau)|.\nonumber
\end{eqnarray}
The last integral corresponds to the one-dimensional
non-relativistic quantum system with the Lagrangian
\begin{eqnarray*}
&& L={\dot{\rho}^2\over4}-{\alpha\over8}|\rho|
\end{eqnarray*}
for which the Hamiltonian reads
\begin{eqnarray}
\label{Ham} && H=p^2+{\alpha\over8}|\rho|
\end{eqnarray}
The  Schr\"{o}dinger equation looks as
\begin{eqnarray}
\label{lineq} &&\left[-{d^2\over
d\rho^2}+{\alpha\over8}|\rho|\right]\Psi(\rho)={\cal E}\Psi(\rho)
\end{eqnarray}
Let us introduce
$$\rho={2v\over\alpha^{{1\over3}}},~~~~~~~~{\cal
E}={\alpha^{{2\over3}}\over4}\epsilon$$ then  $\epsilon$  is the
eigenvalue of the equation
\begin{eqnarray*}
&&\left[-{d^2\over dv^2}+v\right]Y(v,\epsilon)=\epsilon
Y(v,\epsilon),~~~~~~~~~v\in[0,\infty).
\end{eqnarray*}

The non-normalized solution of the equation looks like
\begin{eqnarray*}
&&Y(v,\epsilon)=\left\{\begin{array}{ll}
\pi\sqrt{{\varepsilon-v\over3}}\left[J_{{1\over3}}\left({2\over3}(\varepsilon-v)^{{3\over2}}\right)+
J_{-{1\over3}}\left({2\over3}(\varepsilon-v)^{{3\over2}}\right)\right],
& v<\varepsilon \\
&\\
\sqrt{v-\varepsilon}K_{{1\over3}}\left({2\over3}(v-\varepsilon)^{{3\over2}}\right),
& v>\varepsilon \\
\end{array}\right.
\end{eqnarray*}
The spectrum  is defined by the equations
\begin{eqnarray}
&&\left.{d\over dv}Y(v,\epsilon_{2n})\right|_{v=0}=0~~~~~~~{\rm
for~even~states}~~~~n\to 2n\\
&&Y(0,\epsilon_{2n+1})=0~~~~~~~{\rm for~odd~states}~~~~n\to
2n+1.\nonumber
\end{eqnarray}
The even eigenfunctions $\Phi_{2n}(v)=Y(v,\epsilon_{2n})$ should
have $n$ zeros  and the odd eigenfunctions
$\Phi_{2n+1}(v)=Y(v,\epsilon_{2n+1})$ should have one zero for $v=0$
and $n$ zeros for $0<v<\infty$.

The wave functions are
\begin{eqnarray*}
&&\Phi_n(\rho)=\sqrt{{\alpha^{{1\over3}}\over4N_n}}
Y\left({\alpha^{{1\over3}}\over2}\rho,\epsilon_n\right),~~~~~~\rho\in(-\infty,\infty)
\end{eqnarray*}
The form-factors are defined like
\begin{eqnarray*}
&&  A_n\left({q\over2}\right)=\int\limits_{-\infty}^\infty
d\rho~ e^{i{\alpha q\over2}\rho}~\Phi_0(\rho)\Phi_n(\rho)\\
&&= {\alpha^{{1\over3}}\over4
\sqrt{N_0N_n}}\int\limits_{-\infty}^\infty d\rho~ e^{i{\alpha
q\over2}\rho}~Y\left({\alpha^{{1\over3}}\over2}\rho,\epsilon_0\right)
Y\left({\alpha^{{1\over3}}\over2}\rho,\epsilon_n\right)\\
&&=\left\{\begin{array}{l}
{1\over\sqrt{N_0N_{2n}}}\int\limits_0^\infty dv~
\cos\left(\alpha^{{2\over3}}q v\right)
~Y\left(v,\epsilon_0\right)Y\left(v,\epsilon_{2n}\right)\\
\\
{i\over\sqrt{N_0N_{2n+1}}}\int\limits_0^\infty dv~
\sin\left(\alpha^{{2\over3}}q v\right)
~Y\left(v,\epsilon_0\right)Y\left(v,\epsilon_{2n+1}\right)\\
\end{array}\right.\\
&&={\cal A}_n\left(\alpha^{{2\over3}}q\right),
\end{eqnarray*}
with the symmetry condition
\begin{eqnarray*}
&&{\cal A}_n\left(-\alpha^{{2\over3}}q\right)=(-1)^n{\cal
A}_n\left(\alpha^{{2\over3}}q\right),
\end{eqnarray*}

For $\alpha\ll1$ form-factors behave like
\begin{eqnarray*}
&&{\cal
A}_0\left(\alpha^{{2\over3}}q\right)=1-\alpha^{{4\over3}}~a_0,~~~~~a_0=\int\limits_0^\infty
dv{v^2\over2}~{Y^2\left(v,\epsilon_0\right)\over N_0}=0.374939,\\
&&{\cal A}_1\left(\alpha^{{2\over3}}q\right)= i\alpha^{{2\over3}}q
~a_1,~~~~~a_1=\int\limits_0^\infty dvv~
{Y\left(v,\epsilon_0\right)Y\left(v,\epsilon_1\right)\over\sqrt{N_0N_1}}=0.862863,\\
&&{\cal A}_2\left(\alpha^{{2\over3}}q\right)= \alpha^{{4\over3}}q^2
~a_2,~~~~~a_2=\int\limits_0^\infty dv{v^2\over2}~
{Y\left(v,\epsilon_0\right)Y\left(v,\epsilon_2\right)\over\sqrt{N_0N_2}}=0.569709,\\
&&{\cal A}_3\left(\alpha^{{2\over3}}q\right)= i\alpha^{{2\over3}}q
~a_3,~~~~~a_3=\int\limits_0^\infty dvv~
{Y\left(v,\epsilon_0\right)Y\left(v,\epsilon_3\right)\over\sqrt{N_0N_3}}=-0.0685378,\\
\end{eqnarray*}

The next integral  for $t\to\infty$  behaves like
\begin{eqnarray*}
&&{\cal W}(t)\\
&&=\int{D\rho\over NC} e^{-\int\limits_{0}^td\tau
\left[{1\over4}\dot{\rho}^2(\tau)-U(\rho(\tau))
\right]}\int\limits_0^t
d\tau_1\int\limits_0^{\tau_1}d\tau_2~H(|\tau_1-\tau_2|)
e^{i{\alpha q\over2}(\rho(\tau_1)+\rho(\tau_2))}\\
&&=\int\limits_0^t d\tau_1\int\limits_0^{\tau_1}d\tau_2~H(|\tau_1-\tau_2|)\\
&&\cdot{1\over N} \int d\rho_1 \int d\rho_2~
G_{t-\tau_1}(0,\rho_1)e^{i{\alpha
q\over2}\rho_1}G_{\tau_1-\tau_2}(\rho_1,\rho_2)
e^{i{\alpha q\over2}\rho_2}G_{\tau_2}(\rho_2,0)\\
&&\Rightarrow\int\limits_0^t
d\tau_1\int\limits_0^{\tau_1}d\tau_2~H(|\tau_1-\tau_2|)
\sum\limits_{\kappa} e^{-(\tau_1-\tau_2)(E_{\kappa}-E_{0})}{\cal
A}_{\kappa}\left(\alpha^{{2\over3}}q\right){\cal
A}^*_{\kappa}\left(-\alpha^{{2\over3}}q\right) =t~{\cal W}_0
\end{eqnarray*}
where
\begin{eqnarray*}
&&{\cal W}_0=\int\limits_0^\infty d\tau~H(\tau) \sum\limits_{\kappa}
e^{-\tau(E_{\kappa}-E_{0})}{\cal
A}_{\kappa}\left(\alpha^{{2\over3}}q\right){\cal
A}^*_{\kappa}\left(-\alpha^{{2\over3}}q\right) \\
\end{eqnarray*}
Thus, for $t\to\infty$ one can write
\begin{eqnarray}
&&\left\langle e^{i{\alpha
q\over2}(\rho(\tau_1)+\rho(\tau_2))}\right\rangle_\rho
=\sum\limits_{\kappa}(-1)^\kappa
e^{-{1\over4}|\tau_1-\tau_2|\alpha^{{2\over3}}(\epsilon_{\kappa}-\epsilon_{0})}\left|{\cal
A}_{\kappa}\left(\alpha^{{2\over3}}q\right)\right|^2
\end{eqnarray}

\newpage

\begin{center}
Table 1. QM and QFT.
\end{center}
\begin{center}
\begin{tabular}{|c|c|}
\hline
Quantum Mechanics & Quantum Field Theory \\
\hline
\multicolumn{2}{|c|}{}\\
\multicolumn{2}{|c|}{$H=H_0+gH_I$}\\
\multicolumn{2}{|c|}{}\\
\hline
$(H_0+gH_I)\Psi_E=E\Psi_E$ &$H_0\Psi^{(0)}_E=E\Psi^{(0)}_E$\\
$\left\{\Psi_E\right\}=\left\{free~particles\right\}\bigoplus\left\{bound~states\right\}$&
Fock space$=\left\{free~particles\right\}$\\
& $H_I$ is not operator on Fock space\\
\hline &\\
$\Psi(t)=e^{-iH(t-t_0)}\Psi(t_0)$ &
$S=\lim_{t_0\to-\infty}^{t\to+\infty}e^{iH_0t}e^{-iH(t-t_0)}e^{-iH_0t_0}$\\
Development in time& $\bigoplus$ renormalization\\
& $\Longrightarrow$ $S$ is operator on Fock space\\
& Relation between  asymptotically free states\\
\hline
intermediate particles {\bf ON} mass shell& intermediate particles {\bf OUT OF} mass shell\\
\hline
\multicolumn{2}{|c|}{}\\
\multicolumn{2}{|c|}{Relativistic corrections}\\
\multicolumn{2}{|c|}{}\\
\hline interaction is transmitted instantly & retarded interaction
\\
\hline
Small corrections to $H$& $ A_{in\to out}(p_1,p_2;k_1,k_2)\sim{1\over M^2-(p_1+p_2)^2}$\\
Elimination {\bf TIME} out of Hamiltonian &\\
Effective theories& $\left\langle0|{\bf J}(x){\bf J}(0)|0\right\rangle_0\sim e^{-M|x|},~~~|x|\to\infty$\\
\hline
Breit potential & Bethe-Salpeter equation\\
Nonrelativistic QED & \\
\hline
classification of states ${\bf R}^3$& classification of states ${\bf R}^4={\bf R}^3\bigotimes R^1_t$\\
& "Time" excitations $\Rightarrow$ abnormal states\\
\hline
\end{tabular}
\end{center}

\newpage

\begin{center}
Table 2. Quantum numbers of relativistic currents

\vspace{.5cm}

\begin{tabular}{|c|c|c|c|c|c|c|}
\hline
&&&&&&\\
$J$ &  $(\overline{\psi}O_J\psi)$   & $S$ & $L$  & $J$ &
$P=(-1)^{1+L}$ & $J^P$ \\
&&&&&&\\
\hline\hline
&&&&&&\\
$S$ &  $(\overline{\psi}\psi)~\Longrightarrow~
(\mbox{\boldmath$\sigma k$})$   & $1$ & $1$  & $0$ &$+1$ & $0^+$ \\
&&&&&&\\
\hline
&&&&&&\\
& $(\overline{\psi}\gamma_0\psi)~\Longrightarrow~
(\mbox{\boldmath$\sigma k$})$   & $1$ & $1$  & $0$ & $+1$&$0^+$\\
$V$&&&&&&\\
& $(\overline{\psi}i\mbox{\boldmath$\gamma$}\psi)~\Longrightarrow~
\mbox{\boldmath$\sigma$}$   & $1$ & $0$  & $1$ & $-1$&${\bf 1^-}$\\
&&&&&&\\
\hline
&&&&&&\\
&$(\overline{\psi}\gamma_0\mbox{\boldmath$\gamma$}\psi)~\Longrightarrow~
\mbox{\boldmath$\sigma$}
$   & $1$ & $0$  & $1$ & $-1$&${\bf 1^-}$ \\
$T$&&&&&&\\
&$(\overline{\psi}\sigma_{ij}\psi)~\Longrightarrow~
[\mbox{\boldmath$\sigma\times k$}]$ & $1$ & $1$  & $1$ & $+1$&$1^+$ \\
&&&&&&\\
\hline
&&&&&&\\
& $(\overline{\psi}\gamma_5\gamma_0\psi)~\Longrightarrow~1
$   & $0$ & $0$  & $0$ & $-1$&${\bf 0^-}$ \\
$A$&&&&&&\\
&$(\overline{\psi}\gamma_5\mbox{\boldmath$\gamma$}\psi)~\Longrightarrow~
[\mbox{\boldmath$\sigma\times k$}]$   & $1$ & $1$  & $1$ & $+1$&$1^+$ \\
&&&&&&\\
\hline
&&&&&&\\
$P$ &  $(\overline{\psi}i\gamma_5\psi)~\Longrightarrow~1$
& $0$ & $0$  & $0$ & $-1$&${\bf 0^-}$ \\
&&&&&&\\
\hline
\end{tabular}
\end{center}

\newpage

\begin{center}
Table 3. Functions $\Sigma_\Gamma^{(0)}$ and $\Sigma_\Gamma(k)$
\end{center}

\begin{center}
\begin{tabular}{|c|c|c|c|}
\hline
&&&\\
$J^P$& ${\Gamma}$&$\Sigma_\Gamma^{(0)}$ & $\Delta_\Gamma(k)$\\
&&&\\
\hline\hline
&&&\\
$P(0^-)$& $i\gamma_5\cos\theta_P+\gamma_5\gamma_0\sin\theta_P$ & $-2$&$-2\cdot{{\bf k}^2\over{\bf k}^2+k_4^2}$\\
&&&\\
\hline
&&&\\
$V(1^-)$ & $\gamma_j\cos\theta_V+ i\gamma_0\gamma_j\sin\theta_V$
&$-2\delta_{ij}$ & ${2\over3}\cdot{{\bf k}^2\over
{\bf k}^2+k_4^2}$\\
&&&\\
\hline
&&&\\
$S(0^+)$ & $I \cos\theta+\gamma_0\sin\theta$ & 0 & 0\\
&&&\\
\hline
&&&\\
$A((1^+)$& $\gamma_5\mbox{\boldmath$\gamma$}\cos\theta+
i[\mbox{\boldmath$\gamma$}\times\mbox{\boldmath$\gamma$}]\sin\theta$ &0 & 0\\
&&&\\
\hline
\end{tabular}
\end{center}

\begin{center}
Table 4. The function $\Delta_{n\ell0}(\alpha)$
\end{center}
\begin{center}
\begin{tabular}{|c||c||c|c||c|c|c|}
\hline $(n\ell)$& (10) & (20) & (21)& (30) &
(31) & (32) \\
\hline $\Delta_{n\ell0}(0)$
&1. & 0.0987 & 0.0987 &0.0283& 0.0307 & 0.00244 \\
$\Delta_{n\ell0}(\alpha)$
&0.9999 & 0.09868& 0.09864 &0.02829 & 0.03071 & 0.002438 \\
$\Delta_{n\ell1}(\alpha)$ &0.96707 & 0.09387& 0.09617& 0.2683&0.002986& 0.00238 \\
$\Delta_{n\ell2}(\alpha)$ &0.01421& 0.002089&0.001173&0.0006319&0.0004049 & 0.00002632 \\
\hline
\end{tabular}
\end{center}

\begin{center}
Table 5. Norm $N_n$
\end{center}
\begin{center}
\begin{tabular}{|c|c|c|}
\hline &&\\
$n$ & $\epsilon_n$ & $N_n=\int\limits_0^\infty dv~\Phi_n^2(v)$ \\
&&\\ \hline
0 & 1.0188 & 8.655\\
1 & 2.3381 & 14.558\\
2 & 3.2482 & 16.886\\
3 & 4.0879 & 19.097\\
4 & 4.8201 & 20.652\\
\hline
\end{tabular}
\end{center}

\end{document}